\documentclass[useAMS] {mn2e}

\usepackage{graphicx}
\usepackage{epstopdf}
\usepackage{color}

\title[Evolution of the GDR out to $z \sim 0.5$]
	{Evolution of the Red Sequence Giant to Dwarf Ratio in Galaxy Clusters out to $z \sim$ 0.5}

\author[Bildfell et al.] 
{\parbox{\textwidth}{C.~Bildfell$^{1}$,\thanks{E-mail: \texttt{bildfell@uvic.ca}} 
H.~Hoekstra$^{2}$, 
A.~Babul$^{1}$, 
D.~Sand$^{3}$, 
M.~Graham$^{3}$, 
J.~Willis$^{1}$, 
S.~Urquhart$^{1}$, 
A.~Mahdavi$^{4}$, 
C.~Pritchet$^{1}$, 
D.~Zaritsky$^{5}$, 
J.~Franse$^{2}$, 
P.~Langelaan$^{2}$}\vspace{0.4cm}\\
\parbox{\textwidth}{$^{1}$Department of Physics \& Astronomy, University of Victoria, Victoria, BC V8P 1A1, Canada\\
$^{2}$Leiden Observatory, Leiden University, P.O. Box 9513, NL-2300 RA Leiden, Netherlands\\
$^{3}$Las Cumbres Observatory Global Telescope Network, 6740 Cartona Drive, Suite 102, Santa Barbara, CA 93117, USA\\
$^{4}$Department of Physics \& Astronomy, San Francisco State University, San Francisco, CA 94131, USA\\
$^{5}$Steward Observatory, University of Arizona, Tucson AZ 85721, USA}}

\date{\today}

\begin{document}

\maketitle

\begin{abstract}
We analyze deep $g'$ and $r'$ band data of 97 galaxy clusters imaged with MegaCam on the Canada-France-Hawaii telescope.  We compute the number of luminous (giant) and faint (dwarf) galaxies using criteria based on the definitions of de Lucia et al. (2007).  Due to excellent image quality and uniformity of the data and analysis, we probe the giant-to-dwarf ratio (GDR) out to $z\sim0.55$.  With X-ray temperature ($T_x$) information for the majority of our clusters, we constrain, for the first time, the $T_x$-corrected giant and dwarf evolution separately.  Our measurements support an evolving GDR over the redshift range $0.05 \leqslant z \leqslant 0.55$.  We show that modifying the $(g'-r')$, $m_{r'}$ and $K$-correction used to define dwarf and giant selection do not alter the conclusion regarding the presence of evolution.  We parameterize the GDR evolution using a linear function of redshift ($GDR = \alpha z + \beta$) with a best fit slope of $\alpha = 0.88 \pm 0.15$ and normalization $\beta = 0.44 \pm 0.03$.  Contrary to claims of a large intrinsic scatter, we find that the GDR data can be fully accounted for using observational errors alone.  Consistently, we find no evidence for a correlation between GDR and cluster mass (via $T_x$ or weak lensing).  Lastly, the data suggest that the evolution of the GDR at $z < 0.2$ is driven primarily by dry merging of the massive giant galaxies, which when considered with previous results at higher redshift, suggests a change in the dominant mechanism that mediates the GDR.
\end{abstract}


\section{Introduction \label{introduction}}

Non-starforming galaxies in clusters exhibit a tight sequence in colour-magnitude space known as the red sequence (eg. Visvanathan 1978; Bower et al. 1992), which is in place as early as $z \sim 1$ (eg. Bell et al. 2004, Mei et al. 2006).  Less understood is how the number of red sequence galaxies and their distribution with stellar mass/absolute magnitude evolves with cosmic time.  Understanding the assembly history of the red sequence is critical as these galaxies represent the bulk of the stellar mass in clusters at the present epoch.  A useful metric for probing this assembly history is the ratio of the number of luminous red sequence galaxies (giants) to the number of faint red sequence galaxies (dwarfs).  This is essentially a non-parametric representation of the luminosity function using only 2 bins and is commonly referred to as the Giant-to-Dwarf ratio (GDR).

There is an ongoing debate regarding the presence or absence of evolution in the GDR and/or its reciprocal the Dwarf-to-Giant ratio (DGR).  De Lucia et al. (2007) for instance, finds significant evolution in the GDR, amounting to a change from GDR$\sim0.45$ to GDR$\sim0.95$ over the redshift range $0.4 < z < 0.8$.  The analysis of low redshift ($0.08 < z < 0.19$) cluster data by Capozzi et al. (2010) supports the extrapolation of this trend to lower redshift.  Stott et al. (2007) and Gilbank et al. (2008) also find strong evolution in the DGR using somewhat brighter absolute magnitude limits for the definition of dwarfs.  In contrast, Crawford et al. (2009) look at 59 clusters at redshift $0 < z < 0.5$ and, though they do not measure the GDR directly, conclude that there is little evolution in the faint-end slope of the luminosity function.  Andreon (2008), in an analysis of 28 clusters in the redshift range $0 < z < 1.3$, asserts that the DGR is consistent with no evolution.  More specifically, he states that the strong trend observed by de Lucia et al. (2007) can be ruled out.  Andreon (2008) discusses some of the differences between his analysis and those of de Lucia et al. (2007) and Gilbank et al. (2008), arguing that their selection of filters that do not bracket the 4000 angstrom break over the entire range of redshifts introduces systematic errors that mimic evolution.  Furthermore, he argues that in the work of Stott et al. (2007) and de Lucia et al. (2007) the use of control fields taken in separate filter sets than the cluster fields causes significant systematics related to the interloper-removal process.

The selection of target clusters may also be important.  The Andreon (2008) clusters are all spectroscopically confirmed and identified in X-rays, while those used in de Lucia et al. (2007) are not and may contain line-of-sight superpositions at the low-mass end.

Perhaps the most significant difference between Andreon (2008) and others is the use of a Bayesian approach to fit the luminosity function of red sequence galaxies, which is then converted to a DGR to facilitate comparison with the literature.  It is stated that this is a superior method because it uses all of the available data, particularly at the faint-end of the luminosity function, beyond the limits of the de Lucia et al. (2007) definition of a dwarf galaxy (the most common definition).  We suspect however, that there are distinct disadvantages to the Andreon (2008) type approach because in forcing the parametric form of the luminosity function one is introducing an undesirable model dependence and it may therefore be incorrect to compare with non-parametric results.  Moreover, a procedure that takes all available data and allows the fitted domain to vary from cluster to cluster potentially creates a systematic bias.  This is because the clusters that are sampled to the faintest absolute magnitudes will have more stringent constraints on their faint end slope than those that are truncated at a brighter limit.

With so many subtle differences between the various ways that the GDR has been measured in the literature it may be no surprise that there appears to be some disagreement.  Our analysis, however, is carried out on observations taken in a single filter pair, from a single telescope/instrument (including control fields), and with a sample of exclusively X-ray-confirmed clusters.  These attributes allow us to avoid many of the criticisms mentioned above.  Moreover, we calculate the effect of many potential sources of systematics ($K$-correction, projected distance from the brightest cluster galaxy (BCG), colour selection method, colour error bias and the presence of large scale structure) to estimate their role in the apparent disagreements in the literature.  When coupled with the sheer number of clusters in our sample, this survey provides the most accurate and robust measure of the GDR over this redshift range to date.

If indeed there is an evolving GDR in galaxy clusters, it becomes interesting to investigate the mechanism(s) responsible for driving that evolution.  For instance, gas-poor (dry) merging of galaxies at the bright end of the red sequence could lead to a depletion of the giant population.  In competition with this process, star-forming galaxies move onto the red sequence through the quenching of ongoing star formation and the resulting colour transformation.  However, the observed downsizing behavior of stellar mass assembly (Bell et al. 2004, Juneau et al. 2005, Bundy et al. 2005, Faber et al. 2007, Scarlata et al. 2007) suggests that this effect may become less important with decreasing redshift as the specific star formation rate at a fixed galaxy mass declines.  While these mechanisms are complementary in that they act to reduce the number of giants with respect to dwarfs, their relative importance as a function of look-back time is poorly constrained.  The detailed behaviour of these mechanisms is likely an important benchmark for galaxy formation models.  For example, Gilbank \& Balogh (2008) show that the Galform model of Bower et al. (2006) is unable to reproduce both the observed DGR evolution and the DGR difference between cluster and field populations.
  
As mentioned above, all of our clusters are confirmed in X-rays and the majority of them have X-ray temperature measurements, which can be used as independent mass proxies.  This information is necessary for disentangling any cluster mass dependence and allows us, for the first time, to properly probe the mechanism(s) that drive GDR evolution at these redshifts.

The layout of the paper is as follows.  In \S \ref{data} we discuss the data, along with some tests for completeness and data quality to ensure that our measurements are not significantly biased by systematics.  In \S \ref{analysis} we describe the data reduction and analysis procedure used to measure the red sequence and extract from it the GDR.  Following this, \S \ref{dgr} discusses the results of our GDR analysis along with tests of how varying the selection method can affect the GDR measurement and a detailed comparison with previous results from the literature.  In \S \ref{clusterprops} we discuss the potential dependence of the GDR on cluster mass as implied by the X-ray temperature.  An investigation of the mechanisms responsible for GDR evolution is in \S \ref{driver}.  Finally, \S \ref{conclusions} contains our conclusions.  Throughout this paper we assume a cosmology parameterized by $\Omega_m=0.3$, $\Lambda=0.7$ and $H_0=70$km/s/Mpc.

  \begin{figure}
    \centering
    \includegraphics[width=3.5in]{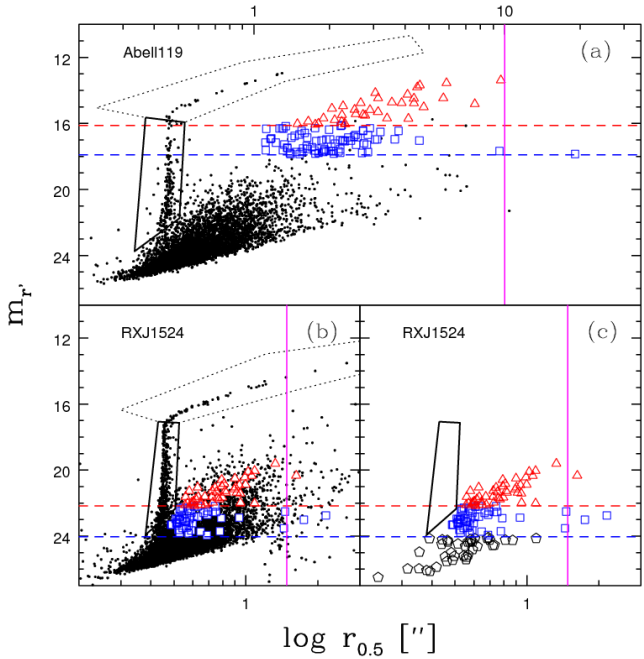}
    \caption{Half-light radius vs apparent magnitude in the $r'$ band for the clusters Abell119 at $z=0.05$(a) and RXJ1524 at $z=0.516$ (b \& c).  The black polygons illustrate our star removal procedure.  The red triangles and blue squares represent all of the giant and dwarf galaxies respectively (see \S \ref{dgr} for details of the selection procedure).  For clarity we show only 5\% of regular objects but 100\% of the selected giants and dwarfs.  The $m_{r'}$ giant and dwarf selection limits are shown by the red and blue dashed lines.  The solid vertical magenta line shows the equivalent size of a $z=0.05$ galaxy measuring 10" after taking into account the PSF and the change in angular diameter distance.  The (c) panel shows only the giants, dwarfs and fainter galaxies (black pentagons) that also matched the red sequence colour selection (see \S \ref{dgr}).  The large size of these fainter galaxies with respect to the stellar distribution indicates a clear separation of stars and red sequence galaxies based on our selection criteria.}
    \label{fig:starcut}
 \end{figure}

\begin{figure}
    \centering
    \includegraphics[width=3.5in]{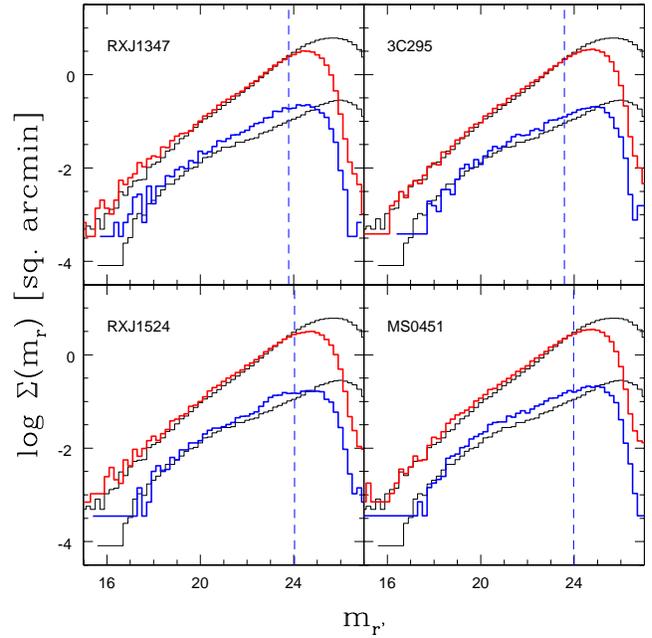}
    \caption{Histograms showing the surface-density of galaxies per sq. arcmin for 4 of the highest redshift clusters in our sample.  The thick lines in each panel show the galaxies detected within a wide (red) and narrow (blue) colour slice (see text).  The corresponding thin black lines show the surface-density of galaxies detected in the CFHTLS Deep images using the same selection criteria.  The blue vertical dashed line shows the location of the faint-end magnitude limit used to select dwarf galaxies (see \S \ref{dgr} for details).  The data are complete up to this limit for the full redshift range investigated.}
    \label{fig:complete_c4}
 \end{figure}

\section{Data} \label{data}

The data for this paper are obtained using the MegaCam\footnote{Based on observations obtained with MegaPrime/MegaCam, a joint project of CFHT and CEA/DAPNIA, at the Canada-France-Hawaii Telescope (CFHT) which is operated by the National Research Council (NRC) of Canada, the Institut National des Science de l'Univers of the Centre National de la Recherche Scientifique (CNRS) of France, and the University of Hawaii.} wide-field imager on the Canada-France-Hawaii Telescope (CFHT) with the $g'$ and $r'$ optical filter set.  The cluster sample is comprised of 101 clusters with redshift $0.05 < z < 0.55$ and is assembled from 3 subsamples: 58 clusters from the Multi-Epoch Nearby Cluster Survey (MENeaCS), 32 clusters from the Canadian Cluster Comparison Project (CCCP) and 11 additional clusters from the CFHT archive.  The MegaCam field is nearly square, subtending approximately 3450 arcseconds on a side.  With the clusters centered in the field of view these data cover projected distances out to 1.49 Mpc from the cluster center at the lowest redshift and 11.05 Mpc at the highest redshift.

To clarify the characteristics of our sample, for the purpose of comparison with other studies, we discuss here some of the selection strategies employed in the sub-samples.  The MENeaCS clusters are selected from the BAX X-ray galaxy cluster database\footnote{http://bax.ast.obs-mip.fr/} and represent all clusters in the database that are observable from CFHT, with redshift $0.05 < z < 0.15$ and X-ray luminosity $L_x > 2 \times 10^{44}$ erg s$^{-1}$.  There are 55 clusters satisfying these criteria, to which we add 3 clusters with the slightly relaxed criteria, $L_x > 1.5 \times 10^{44}$ erg s$^{-1}$ and $0.044 < z < 0.15$ for a total of 58.  The CCCP consists of 50 clusters in the redshift range $0.15 < z < 0.55$, selected from the cluster catalog of Horner (2001).  The CCCP clusters all have X-ray temperatures $T_x > 3$ keV and are selected such that they fully sample the observed scatter in the $L_x$ vs $T_x$ plane.  For the purpose of maintaining a uniform observational configuration, we restrict the current analysis to the 32 CCCP clusters observed with MegaCam $g'$ and $r'$.  Lastly, as mentioned above, we add 11 clusters from the CFHT archive that are also observed with this instrumental configuration.

All of our data are initially calibrated with CFHT's Elixir software pipeline.  The Elixir analysis includes a photometric zeropoint that is determined for each individual run. To account for non-photometric conditions, the magnitudes of objects in common between exposures are compared.  For the CCCP photometric conditions were requested and the comparison of exposures indicates that the uncertainty in the zeropoints are $< 0.05$ magnitudes, which implies an impact on the colour that is smaller. The MENeaCS data are obtained over many queue runs, most of which were during photometric conditions. The results for clusters that overlap with SDSS show agreement within $< 0.04$ magnitudes.  The archival data were reprocessed using the same technique of comparing magnitudes of objects common to multiple exposures and similarly, we find that the uncertainty on individual zeropoints is $< 0.05$ magnitudes.

From the combined sample of 101 clusters, we remove 4 from the analysis because of their low richness (see \S \ref{bgcolcor}) leaving a total of 97.  Basic information for our targets is listed in Table \ref{tab:targets}.  Further details on the CCCP survey will be presented in a future paper (Hoekstra in prep.), while a more detailed description of the MENeaCS survey can be found in Sand et al. (2012).  The largest seeing disc of our images has a FWHM of 1.0" with the majority being significantly smaller.  For background determination, we make use of the CFHT Legacy Survey (CFHTLS) Deep images.  These data are processed and hosted by S. Gwyn\footnote{http://www1.cadc-ccda.hia-iha.nrc-cnrc.gc.ca/community/CFHTLS-SG/docs/cfhtlsDeep.html} (see Gwyn 2008 for details) and consist of four fields that are not targeted at, nor dominated by, individual galaxy clusters.  The CFHTLS Deep fields are observed using the same instrument and filters as our targeted observations but to greater depth.  As such, they provide excellent control fields for statistical background subtraction (discussed further in \S \ref{dgr}).

\begin{table*}
\caption{Table of clusters used in our sample.  RA ($\alpha$) and DEC ($\delta$) give the coordinates of the BCG.  Cluster identifiers marked with a * are excluded from the final analysis due to low richness (see \S \ref{bgcolcor}).}
\begin{minipage}[b]{0.45\linewidth}\centering
\begin{tabular}{| l | r | r | r |}
\hline
Name & \multicolumn{1}{|c|}{$z$} & \multicolumn{1}{|c|}{$\alpha$} & \multicolumn{1}{|c|}{$\delta$}\\
            &         & hh:mm:ss & dd:mm:ss\\
\hline
Abell119 &  0.044 &   00:56:16.10 &  -01:15:19.0\\
MKW3S &  0.045 &  15:21:51.84 &   07:42:31.9\\
Abell780 &  0.054 &   09:18:05.65 & -12:05:43.5\\
Abell754 &  0.054 &   09:08:32.37 &  -09:37:47.2\\
Abell85 &  0.055 &   00:41:50.45 &  -09:18:11.1\\
Abell2319 &  0.056 &  19:21:10.00 &  43:56:44.5\\
Abell133 &  0.057 &   01:02:41.71 & -21:52:55.2\\
Abell1991 &  0.059 &  14:54:31.48 &  18:38:33.4\\
Abell1781* &  0.062 &  13:44:52.54 &  29:46:15.6\\
Abell1795 &  0.062 &  13:48:52.49 &  26:35:34.8\\
Abell553 &  0.066 &   06:12:41.09 &  48:35:44.6\\
Abell644 &  0.070 &   08:17:25.61 &  -07:30:45.0\\
Abell399 &  0.072 &   02:57:53.09 &  13:01:51.3\\
Abell2065 &  0.073 & 15:22:29.17 & 27:42:27.8\\
Abell401 &  0.074 &   02:58:57.78 &  13:34:58.3\\
ZwCl1215 &  0.075 &  12:17:41.13 &   03:39:21.2\\
Abell2670 &  0.076 &  23:54:13.67 & -10:25:08.2\\
Abell2029 &  0.077 &  15:10:56.09 &   05:44:41.4\\
Abell2495 &  0.078 &  22:50:19.71 &  10:54:12.8\\
RXSJ2344-04 &  0.079 &  23:44:18.20 &  -04:22:48.8\\
ZwCl0628 &  0.081 &   06:31:22.69 &  25:01:06.8\\
Abell2033 &  0.082 &  15:11:26.51 &  06:20:56.8\\
Abell1650 &  0.084 &  12:58:41.49 &  -01:45:41.0\\
Abell1651 &  0.085 &  12:59:22.49 &  -04:11:45.7\\
Abell2420 &  0.085 &  22:10:18.76 & -12:10:13.9\\
Abell2597 &  0.085 &  23:25:19.72 & -12:07:26.7\\
Abell763* &  0.085 &   09:12:35.18 &  16:00:01.0\\
Abell478 &  0.088 &   04:13:25.27 &  10:27:55.1\\
Abell2440 &  0.091 &  22:23:56.92 &  -01:34:59.5\\
Abell2142 &  0.091 &  15:58:19.99 &  27:14:00.5\\
Abell1927 &  0.095 &  14:31:06.79 &  25:38:01.6\\
Abell21 &  0.095 &   00:20:36.98 &  28:39:33.0\\
Abell2426 &  0.098 &  22:14:31.58 & -10:22:26.1\\
Abell2055 &  0.102 &  15:18:45.72 &   06:13:56.4\\
Abell1285 &  0.106 &  11:30:23.82 & -14:34:52.3\\
Abell7 &  0.106 &   00:11:45.25 &  32:24:56.6\\
Abell2064* &  0.108 &  15:20:52.24 &  48:39:38.7\\
Abell2443 &  0.108 &  22:26:07.92 &  17:21:23.8\\
RXCJ0352+19 &  0.109 &   03:52:58.99 &  19:40:59.8\\
Abell2703 &  0.114 &   00:05:23.95 &  16:13:09.3\\
Abell2069 &  0.116 &  15:24:08.42 &  29:52:55.6\\
Abell1361 &  0.117 &  11:43:39.60 &  46:21:20.7\\
Abell2050 &  0.118 &  15:16:17.92 &   00:05:20.9\\
RXCJ0736+39 &  0.118 &   07:36:38.08 &  39:24:52.8\\
Abell1348 &  0.119 &  11:41:24.19 & -12:16:38.5\\
Abell961 &  0.124 &  10:16:22.86 &  33:38:17.7\\
Abell1033 &  0.126 &  10:31:44.32 &  35:02:29.2\\
Abell2627 &  0.126 &  23:36:42.08 &  23:55:29.5\\
Abell655 &  0.127 &   08:25:29.05 &  47:08:00.9\\
\hline
\end{tabular}
\end{minipage}
\hspace{0.5cm}
\begin{minipage}[b]{0.45\linewidth}
\centering
\begin{tabular}{| l | r | r | r |}
\hline
Name & \multicolumn{1}{|c|}{$z$} & \multicolumn{1}{|c|}{$\alpha$} & \multicolumn{1}{|c|}{$\delta$}\\
            &         & hh:mm:ss & dd:mm:ss\\
\hline
Abell646 &  0.129 &   08:22:09.53 &  47:05:53.3\\
Abell1132 &  0.136 &  10:58:23.65 &  56:47:42.0\\
Abell795 &  0.136 &   09:24:05.29 &  14:10:21.8\\
Abell1068 &  0.138 &  10:40:44.48 &  39:57:11.5\\
Abell1413 &  0.143 &  11:55:18.00 &  23:24:18.1\\
ZwCl1023 &  0.143 &  10:25:57.98 &  12:41:08.7\\
Abell990 &  0.144 &  10:23:39.91 &  49:08:38.8\\
Abell2409 &  0.148 &  22:00:53.49 &  20:58:42.1\\
RXCJ0132-08* &  0.149 &   01:32:41.11 &  -08:04:04.6\\
Abell2204 &  0.150 &  16:32:46.96 &   05:34:33.0\\
Abell545 &  0.154 &   05:32:25.18 & -11:32:39.3\\
Abell2104 &  0.160 &  15:40:07.92 &  -03:18:16.1\\
Abell2259 &  0.160 &  17:20:09.65 &  27:40:08.5\\
Abell1234 &  0.166 &  11:22:29.94 &  21:24:22.2\\
Abell1914 &  0.170 &  14:26:03.88 &  37:49:53.8\\
Abell586 &  0.170 &   07:32:20.30 &  31:38:01.3\\
Abell1246 &  0.190 &  11:23:58.82 & 21:28:50.0\\
MS0440 &  0.190 &   04:43:09.90 &   02:10:19.5\\
Abell115S &  0.200 &   00:56:00.25 &  26:20:32.8\\
Abell115N &  0.200 &   00:55:50.61 &  26:24:37.7\\
Abell2163 &  0.200 &  16:15:48.98 &  -06:08:41.0\\
Abell2261 &  0.200 &  17:22:27.22 &  32:07:57.9\\
Abell520 &  0.200 &   04:54:14.05 &   02:57:10.8\\
Abell223N &  0.210 &   01:38:02.28 & -12:45:19.7\\
Abell223S &  0.210 &   01:37:55.98 & -12:49:09.9\\
Abell1942 &  0.220 &  14:38:21.87 &   03:40:13.5\\
Abell2111 &  0.230 &  15:39:40.50 &  34:25:27.7\\
Abell2125 &  0.246 &  15:41:14.75 &  66:16:03.8\\
Abell1835 &  0.250 &  14:01:02.09 &   02:52:42.9\\
Abell521 &  0.250 &   04:54:06.87 & -10:13:24.4\\
CL1938 &  0.260 &  19:38:18.10 &  54:09:40.4\\
Abell1758W &  0.280 &  13:32:38.41 &  50:33:36.2\\
Abell697 &  0.280 &   08:42:57.56 &  36:21:59.6\\
Abell611 &  0.290 &   08:00:56.82 &  36:03:24.0\\
Abell959 &  0.290 &  10:17:36.00 &  59:34:02.0\\
Abell2537 &  0.300 &  23:08:22.21 &  -02:11:31.5\\
MS1008 &  0.300 &  10:10:32.32 & -12:39:52.6\\
Abell851 &  0.410 &   09:42:57.45 &  46:58:50.1\\
RXJ0856 &  0.411 &   08:56:12.69 &  37:56:15.9\\
RXJ2228 &  0.412 &  22:28:33.70 &  20:37:16.6\\
RXJ1003 &  0.416 &  10:03:04.62 &  32:53:41.4\\
MS1621 &  0.426 &  16:23:35.14 &  26:34:28.3\\
MACS1206 &  0.440 &  12:06:12.14 &  -08:48:03.1\\
CL0910 &  0.440 &   09:13:45.50 &  40:56:28.7\\
RXJ1347 &  0.450 &  13:47:30.64 & -11:45:08.9\\
RXJ1701 &  0.453 &  17:01:23.51 &  64:14:12.0\\
3C295 &  0.460 &  14:11:20.55 &  52:12:10.1\\
RXJ1524 &  0.516 &  15:24:38.37 &   09:57:43.6\\
MS0451 &  0.540 &   04:54:10.83 &  -03:00:51.2\\
\hline
\end{tabular}
\end{minipage}
\label{tab:targets}
\end{table*}

\subsection{Source Extraction} \label{extract}

The data are bias-subtracted and flat fielded using the Elixir software package.  We combine the images and perform background subtraction using SWarp (Bertin et al. 2002).  After data reduction and image stacking we mask obvious image defects and regions around bright stars that could severely bias the photometry.  We then run SExtractor (Bertin \& Arnouts 1996) in dual image mode on each set of $g'$ and $r'$ images using the masks merged with exposure maps as weight images.  To maximize the detection of faint dwarf galaxies, even when they are embedded in the halos of much larger giant galaxies, we adjust some of the SExtractor parameters from their default values.  We briefly discuss here some of these parameters that can affect the number of sources detected in crowded regions and the values we use for our extractions.  The DETECT\_THRESH and ANALYSIS\_THRESH values are set to 1.5 relative to the background variance and the DETECT\_MINAREA is set to 3 pixels.  To enable the detection of faint galaxies near to bright ones in these crowded fields we set the DEBLEND\_MINCONT to $10^{-7}$ and the DEBLEND\_NTHRESH to 64.  There are also several parameters that control the way the background is measured and subtracted from the detected sources and, therefore, these can affect the apparent magnitudes of galaxies especially at the faint end.  For these we use a BACK\_SIZE of 128 pixels and a BACK\_FILTERSIZE of 3 pixels.  Lastly, SExtractor has the option to ``CLEAN" the source catalog of artifacts and false detections.  We find that setting CLEAN $=$Y with a CLEAN\_PARAM$=2.0$ not only removes many of the artifacts and false detections but also increases the number of real objects detected compared to the case of CLEAN$=$N.  Tests comparing these results to those obtained using more relaxed parameters (ie. DEBLEND\_MINCONT$=10^{-4}$, DEBLEND\_NTHRESH=16 and CLEAN=N) show that our modified parameter choices can yield of order 2 additional dwarfs and giants per cluster.

Throughout this paper we use SExtractor MAG\_AUTO for galaxy magnitudes and MAG\_APER for galaxy colours.  The aperture magnitudes are extracted using a 3" diameter aperture, which is several times larger than the worst case seeing disc of the sample (see \S \ref{data}).  All catalogs are corrected for foreground galactic extinction using E(B-V) values from Schlegel et al. (1998) and the extinction curve of Cardelli et al. (1989).

 \begin{figure}
    \centering
    \includegraphics[width=3.2in]{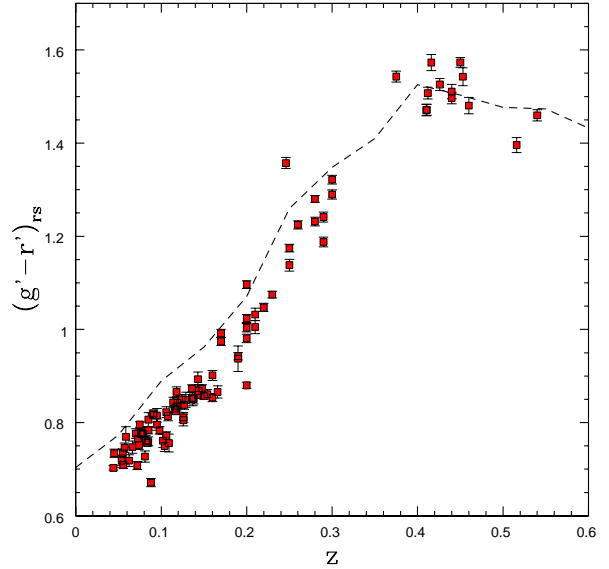}
    \caption{The average observed ($g'-r'$) colour of red sequence galaxies with $-20.82 < M_{r'} < -19.82$ determined using a bootstrap method.  This absolute magnitude bin corresponds to half a mag on either side of the division between dwarfs and giants.  The data are plotted as a function of redshift.  The strong correlation observed is the effect of the $g'$ filter moving across the 4000 $\AA$ break with increasing redshift.  The dashed line represents a simple model of the red sequence (see Appendix \ref{redseqappendix} for details)}
    \label{fig:rsc}
 \end{figure}

 \begin{figure}
    \centering
    \includegraphics[width=3.2in]{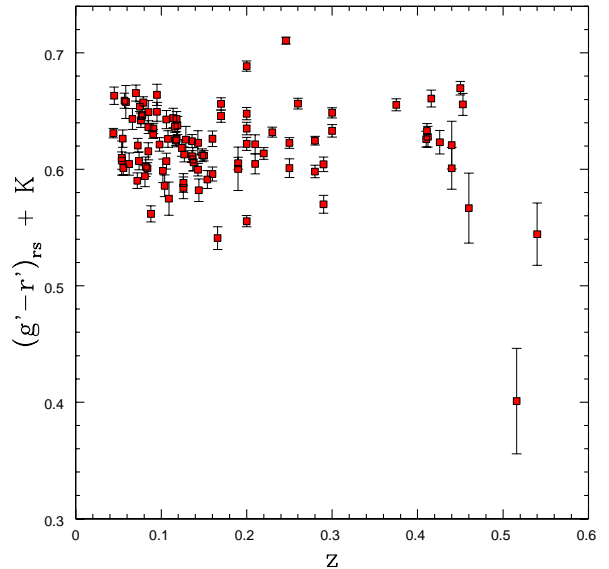}
    \caption{The average K-corrected ($g'-r'$) colour of red sequence galaxies with $-20.82 < M_{r'} < -19.82$ determined using a bootstrap method.  The data are plotted as a function of redshift. It is clear that after $K$-correction the mean ($g'-r'$) colour at the division between dwarf and giant galaxies is roughly constant.}
    \label{fig:rsck}
 \end{figure}

\subsection{Stars \& Extended Galaxies} \label{stars}

The object catalogs discussed in \S \ref{extract} include stars.  To obtain a galaxy catalog we remove the objects that lie along the locus defined by point sources in the half-light radius ($r_{0.5}$) vs apparent $r'$ magnitude ($m_{r'}$) plane.  Figure \ref{fig:starcut} illustrates this process for two clusters in our sample; one at the lowest redshift and one at the highest.  The thick, solid, black box outlines the objects that are identifiable as unsaturated stars while the thin dotted polygon similarly identifies the saturated ones.  At magnitudes fainter than the limit of the solid box it becomes impossible to confidently distinguish point sources from extended ones by this method.  For this reason, we retain all sources fainter than this limit in our galaxy catalogs so as not to remove the faintest galaxies.  We verify, however, that any remaining stellar contamination does not bias our results by examining whether the GDR (as measured in \S \ref{dgr}) changes when excluding clusters at galactic latitude $b<35$ degrees, which it does not.
 
The blue squares and red triangles in Figure \ref{fig:starcut} show the red sequence dwarf and giant galaxies respectively, with the blue and red dashed lines showing the corresponding magnitude limits for their selection.  These limits are adopted from the de Lucia et al. (2007) definitions (dwarfs: $-20.0< M_V < -18.2$ \& giants: $M_V < -20.0$) which are converted to the corresponding $m_{r'}$ values at the appropriate redshift (see \S \ref{dgr} for details of this conversion).  We note the presence of a small number of detections with $r_{0.5} > 1.0$" and $m_{r'} > 24.0$ for some images (eg. RXJ1524).  We have investigated these by eye and conclude that they are all false detections and, because they do not enter the dwarf and giant selection limits, they do not affect the GDR.

To verify that the number of dwarfs or giants is not affected by the detection efficiency of galaxies with relatively low central surface brightness, we examine the most extended galaxies near the detection limit in each cluster.  From visual inspection of $r_{0.5}$ vs $m_{r'}$ (eg. Figure \ref{fig:starcut}), we find that at $z\sim0.05$ there are very few galaxies detected that are larger than 10".  At $z\sim0.05$, after subtracting in quadrature the size of the PSF, we estimate the physical size of the limiting half-light radius to be $r_{0.5}\sim8.5$ kpc.  This limit is then converted back to an angular size at the appropriate redshift, corrected for PSF effects and then overplotted as the solid magenta line in Figure \ref{fig:starcut}.  The analogs of these extended galaxies are well above our detection limit in the high $z$ end of our sample.  We conclude that the evolution of the number of dwarf and giant galaxies is not affected by the detectability of dwarfs with extended surface brightness profiles.  Furthermore the fraction of such galaxies is small in both the dwarf and giant populations.

\subsection{Completeness} \label{complete}

 To investigate the completeness of our galaxy catalogs we compare them with the catalogs obtained from the CFHTLS Deep data (see \S \ref{data}).  We generate Galactic extinction-corrected, star-cleaned catalogs in both $g'$ and $r'$ for each of the 4 Deep fields using the same procedure as described in sections \ref{extract} and \ref{stars}.  We compare the number of galaxies as a function of magnitude $m_{r'}$ in our sample with those in the CFHTLS Deep (control) using both a wide colour window $[-1.0<(g'-r')<4.0]$ and a narrow one $[1.0<(g'-r')<1.2]$.  Figure \ref{fig:complete_c4} shows the surface density of galaxies selected with these criteria for 4 clusters at the upper redshift limit of our sample.  The thick red and blue histograms show the distributions of sources outside a projected radius of 750 kpc in each of our cluster fields for the wide colour window (red) and the narrow window (blue) while the underlying thin black histograms show the corresponding distributions of sources in the CFHTLS Deep survey area.  The vertical blue dashed lines show the location of the faint end limit for dwarf galaxies (see \S \ref{dgr}).  The slopes of the counts distributions are in agreement between our cluster data and the CFHTLS Deep fields.  Some cluster fields show a small excess of objects with respect to the control sample, but this is expected due to enhancement from cluster galaxies that lie further than 750 kpc from the BCG.  Some fields also exhibit a small jump at $m_{r'}\sim 22$ (eg. RXJ1524, MS0451) which corresponds to the apparent magnitude where we can no longer reliably distinguish stars from galaxies.  The clusters that exhibit the largest jumps are those in the galactic plane $b<35^{\circ}$.  As mentioned in \S \ref{stars} we verified that these clusters do not bias our results by repeating our analysis with $b<35^{\circ}$ removed and obtaining consistent measurements of the GDR evolution.  The CFHTLS Deep data is several magnitudes deeper in the $r'$ band than our cluster sample, thus the surface-density begins to fall off in the clusters fields at brighter $m_{r'}$ than in the control fields.  In all cases however, the sharp drop in counts observed in our sample occurs at magnitudes fainter than the faint dwarf limit (vertical dashed line).  We conclude that the GDR measurements presented in \S \ref{dgr} are not affected by incompleteness in the dwarf population.

\section{Analysis} \label{analysis}

In this section we describe the various criteria we use to select galaxies belonging to the dwarf and giant populations within clusters and present some simple tests to show that these selection criteria are robust.

\subsection{Red Sequences} \label{redseq}

With the unambiguous stars removed from our catalogs, we plot the colour-magnitude diagrams (CMD) of the remaining objects and fit their red sequences.  We present an overview of the red sequence fitting results here but for a more detailed description of the fitting procedure we refer the reader to Appendix \ref{redseqappendix}.  Briefly, we use the biweight estimator of Beers et al. (1990) to fit the colour-magnitude relation (CMR), employing a statistical background subtraction method similar to that of Pimbblet et al. (2002). 

To check the quality of our photometry and red sequence fitting procedure, we plot in Figure \ref{fig:rsc} the colour of the red sequence ($(g'-r')_{rs}$) at a fixed $M_{r'}$ versus redshift.  The $(g'-r')_{rs}$ colour shown is calculated using a bootstrap method, finding the average $(g'-r')$ of the red sequence member galaxies lying within $\Delta m_{r'} = \pm 0.5$ mag on either side of the $m_{r'}$ value that separates dwarfs from giants (see \S \ref{dgr} for detail of dwarf and giant selection).  The strong dependence of $(g'-r')_{rs}$ colour on redshift seen in Figure \ref{fig:rsc} is a result of the 4000 angstrom break spectral feature, that is prominent in early type galaxies, shifting across the $g'$ band with increasing redshift.  The errors shown, which are typically smaller than the symbol size, account only for statistical variance and do not include any systematics.  We also show, as a dashed line, the expected $(g'-r')_{rs}$ for our model red sequences (see appendix \ref{redseqappendix}).  The slope of the $(g'-r')_{rs}$ evolution is similar between the observations and the model, with a flattening at $z\sim0.4$.  There is however a significant systematic offset between the two from $0.05 < z < 0.3$, as discussed in appendix \ref{redseqappendix} and is likely a consequence of the overly simplistic star formation history assumed in the model.

\subsection{Magnitude Limits} \label{maglim}

We measure the ratio of luminous (giant) to faint (dwarf) red sequence galaxies (defined as $|\Delta (g'-r')| < 0.2$ from the best fit red sequence) using the absolute magnitude cuts defined by de Lucia et al. (2007) $M_V=[-20.0, -18.2]$ which we transform to Megacam $M_{r'}$ using the single stellar population red sequence model described in appendix \ref{redseqappendix}.  Thus, we define giant galaxies as those with $M_{r'} < -20.32$ and dwarf galaxies as those with $-20.32 < M_{r'} < -18.52$.  Note that these converted limits are identical to those used in Capozzi et al. (2010), who looked at the GDR using similar filters in the Sloan digital sky survey (SDSS).  The excellent agreement with the Capozzi et al. (2010) limits at all redshifts indicates that the difference introduced by comparing across photometric systems is negligible.

At this stage, a choice in methodology must be made before these cuts can be applied to the data.  One can either transform the data to absolute magnitudes or transform the absolute magnitude cuts to apparent magnitudes.  We choose to do the latter because the former method requires a $K$-correction to all of the data, and since a $K$-correction based on only the $g'$ and $r'$ filters becomes degenerate beyond $z\sim0.6$ this would allow additional high-redshift background objects to enter the dwarf and giant samples.  Therefore our dwarf/giant sample selection is performed in observed space by transforming the above-mentioned absolute magnitude cuts following the relation:

\begin{equation}
m_{r'} = M_{r'} + 5 \log(D_l) + 25 - (K + E)
\label{eqn:obscuts}
\end{equation}

\noindent where $D_l$ represents the luminosity distance of the cluster in Mpc and the $K$-and $E$ terms represent the $K$-correction and the evolution correction respectively.  These corrections are described in detail in \S \ref{kpluse}.

\subsection{$K+E$ Correction} \label{kpluse}

We assume that the luminosity evolution correction ($E$) takes the form of $2.5 \times \log(1+z)$, which is appropriate for passively evolving elliptical galaxies on the red sequence.  The $K$-correction is calculated from the Charlot \& Bruzual (CB07) synthetic spectra, assuming a solar metallicity $Z_\odot=0.02$ with a 3\% contribution by stellar mass from ultra-metal poor stars (see appendix \ref{redseqappendix}).  The age of the spectral template is assigned by matching its redshifted $(g'-r')$ colour to the $(g'-r')$ colour of a galaxy on the best fitting red sequence at a given $m_{r'}$.  The appropriate $K$-correction is then calculated and this procedure is repeated for a sample of $m_{r'}$ values along the red sequence.  Inverting equation \ref{eqn:obscuts} we generate the function that describes the $K$-correction as it depends on absolute magnitude.  This is then interpolated for the dwarf and giant absolute magnitude limits described above to obtain the $K$-corrections that are specific to the colour-magnitude coordinates of said limits along the best fit red sequence.

To verify that our $K$-correction is working properly we use a bootstrap method to compute the colour of the red sequence member galaxies in an $m_{r'}$ bin that extends $\pm 0.5$ around the cut in $m_{r'}$ that divides the dwarfs from the giants.  Figure \ref{fig:rsck} shows the redshift dependence of this colour after $K$-correction.  Once the $K$-correction has been accounted for the $(g'-r')$ colour around the dwarf/giant division remains relatively flat.  There is one cluster (RXJ1524) that is a significant outlier, at $z\sim0.5$ with a $K$-corrected $(g'-r')_{rs} \sim 0.4$.  Inspection of the colour-magnitude diagram for this cluster reveals a second red sequence located at slightly bluer colour than that of RXJ1524, which is consistent with being the source of the observed colour bias.  This contamination likely also biases the GDR measurement for this cluster but it does not affect our conclusions regarding the GDR of the ensemble.  The constant colour seen in Figure \ref{fig:rsck} is expected for massive early type galaxies in this redshift range and lends confidence to the $K$-correction procedure.  The direct impact of the $K$-correction on GDR is investigated in \S \ref{selcut}.

\subsection{Giant-to-Dwarf Colour Difference}

 \begin{figure}
    \centering
    \includegraphics[width=3.2in]{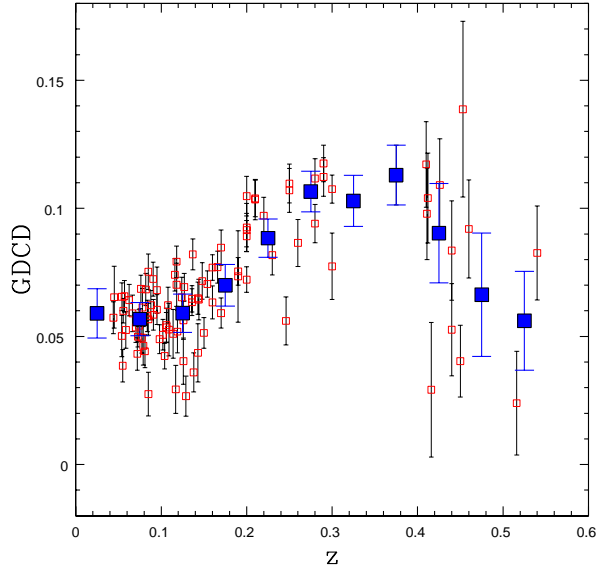}
    \caption{The giant-to-dwarf colour difference (GDCD) as it depends on redshift.  Large symbols represent the error-weighted mean GDCD in each redshift bin.  As a differential measurement, GDCD is unaffected by photometric zeropoint errors, making it a good benchmark for models of the red sequence.}
    \label{fig:gdcd}
 \end{figure}

As a further check, we compute the $(g'-r')$ colour of both dwarf and giant galaxies and examine the evolution of the giant-to-dwarf colour difference (GDCD) as defined by:

\begin{equation}
GDCD_{(g'-r')}=\langle (g'-r')_g \rangle - \langle (g'-r')_d \rangle
\end{equation}

\noindent where $\langle (g'-r')_g \rangle$ and $\langle (g'-r')_d \rangle$ are the mean colours of giant and dwarf galaxies respectively as calculated for individual clusters. Because the GDCD is a relative measurement, it is not affected by systematics that induce a shift in the photometric zero point over the entire image.  Thus, it may provide a useful test of the galaxy properties in cluster formation simulations.  Figure \ref{fig:gdcd} shows the GDCD plotted against redshift.  This quantity is effectively a measure of the red sequence slope as sampled by the individual member galaxies.  From the model red sequences we expect the slope to increase with redshift out to $z\sim0.4$ and then turn over as the ($g'-r'$) filter combination becomes less sensitive to the 4000 angstrom break.  Moreover, this loss of sensitivity to the 4000 angstrom break results in a reduced color separation between red sequence and blue cloud members, increasing the level of red sequence contamination.  Such behavior is reflected in the GDCD, though with a significant scatter.  This highlights the concern that the level of contamination by the blue cloud is a function of redshift and could falsely contribute to the evolution signal in the GDR.  We attempt to correct for this effect in \S \ref{bgcolcor}, but we note here that if it were artificially creating a redshift trend in the GDR, then we would expect to see similar GDR values at both $z=0.1$ and $z=0.5$.  This is not the case.

\section{Giant-to-Dwarf Ratio} \label{dgr}

In this section we discuss issues that deal directly with the GDR measurements.  First, in section \S \ref{bgcolcor} we describe the statistical background subtraction method and corrections for biasing effects introduced by uncertainty in $(g'-r')$ colour.  This is followed in \S \ref{gdrresults} by the results of GDR measurements for individual clusters.  As mentioned in \S \ref{maglim}, we use magnitude limits defined by de Lucia et al. (2007) and convert to $m_{r'}(z)$.  Our criterion for colour selection of dwarf and giant galaxies on the red sequence is based on the offset in $(g'-r')$ from the best fit red sequence ($|\Delta (g'-r')| < 0.2$).  We further require as a spatial criterion that the projected distance from the BCG is $R < 750$ kpc.  Unless stated otherwise, the union of these three criteria define our dwarf and giant selection and are referred to as the default selection method.  Within the literature however, there is no clear consensus on a set of criteria that defines the dwarf and giant populations, owing in part to varying data quality, instrumentation and redshift range of observation.  To facilitate a more direct literature comparison, in \S \ref{selcut} we examine the sensitivity of the GDR to variations in the selection criteria described above.  Lastly, in \S \ref{litcomp}, we compare our GDR values to previous results.

\subsection{Background Subtraction and Colour-Error Correction} \label{bgcolcor}

We apply the selection limits discussed above to our CMDs to find the number of dwarfs and giants in each cluster.  The resulting numbers, however, are contaminated by interlopers.  To correct for this contamination we apply a statistical subtraction method that uses the catalogs generated from the CFHTLS Deep field images (see \S \ref{complete}).  For each cluster, we take the $m_{r'}$ and $(g'-r')$ limits for dwarfs and giants and apply these same limits to the CMD of the merged Deep field catalogs.  We then correct the cluster counts in the following way:

\begin{equation}
n_c = n_o - n_b \frac{A}{A_b}
\label{eqn:bgd}
\end{equation}

\noindent where $n_c$ is the background-corrected number of cluster galaxies, $n_o$ is the observed number of galaxies, $n_b$ is the number of galaxies in the background catalog and $A$ and $A_b$ are the area of the target and background samples respectively.

An additional correction is made to account for the bias introduced by the uncertainty in the observed $(g'-r')$ colour.  For instance, an intrinsically narrow red sequence will be broadened colourwise, which can result in galaxies being scattered outside of the colour selection limits, especially at the highest redshifts.  It is also possible for cluster members in the blue cloud to be observationally scattered into the red sequence, increasing the resulting number of red sequence galaxies and acting in opposition to the previously described effect.  Both are statistically significant effects and failure to account for this observational bias can mimic an evolution in the GDR.

To correct for the colour-error bias, we estimate the number of galaxies observed to be red sequence members as a fraction of the number that are intrinsically on the red sequence (ie. the selection fraction).  We simulate this process by taking the distribution of $m_{r'}$ and $(g'-r')$ and perturbing each value by a normally distributed random number that is scaled to the $1\sigma$ measurement uncertainty on each of these quantities.  We compute the background-subtracted number of dwarfs and giants both before and after this perturbation and repeat the procedure for 1000 realizations.  No perturbation is applied to the background galaxies as their color-magnitude errors are much smaller than those of the target sample.  We assume that the selection fraction ($f_{s}$) can be approximated by $f_{s}\approx n_{rs}/n_{rs}^{\prime}$ where $n_{rs}$ and $n_{rs}^{\prime}$ are the number of red sequence galaxies before and after the perturbation respectively.  This approximation is valid as long as the difference $(n_{rs}-n_{rs}^{\prime}$) is small compared to $n_{rs}$.  For the giants and dwarfs below $z<0.3$ the effect is of order 5\%, beyond $z>0.3$ it grows to a maximum of 20\%.  The appropriate selection fraction is calculated for both the dwarf and giant populations and is compensated for by applying $f_s$ to $n_c$ from equation \ref{eqn:bgd} to compute the total number of galaxies $n_t=n_c/f_s$ respectively.

 \begin{figure}
    \centering
    \includegraphics[width=3.2in]{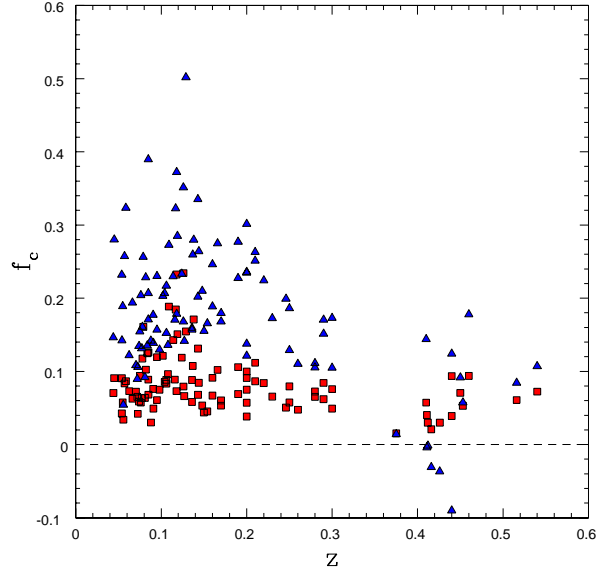}
    \caption{The effect of background subtraction and colour-error correction expressed as a fraction of the total corrected counts for dwarfs (blue triangles) and giants (red squares) plotted against cluster redshift.  The correction fraction becomes slightly negative for the high redshift dwarfs when the colour-error correction begins to dominate over the background subtraction.}
    \label{fig:bgsub}
 \end{figure}

 \begin{figure}
    \centering
    \includegraphics[width=3.2in]{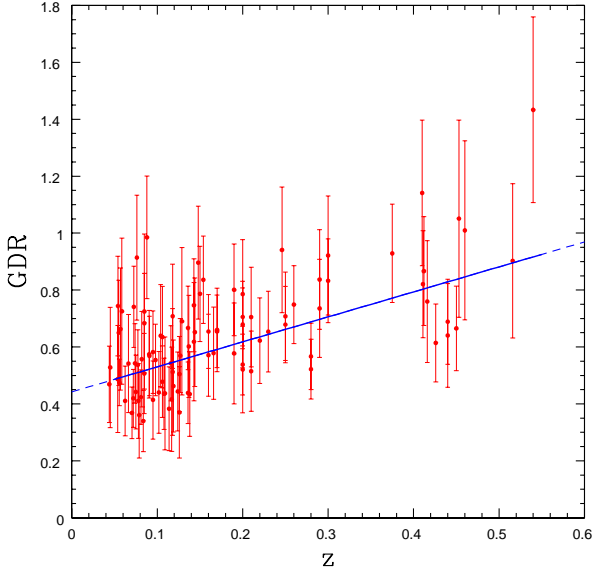}
    \caption{The Giant-to-Dwarf ratio (GDR) plotted against redshift.  Individual clusters are marked by the filled red symbols.  The blue line show the best fit linear redshift dependence.}
    \label{fig:gdr_plain}
 \end{figure}

To assess the importance of background subtraction and the colour-error correction we show in Figure \ref{fig:bgsub} the fraction of galaxies ($f_c$) that must be subtracted to correct for these effects as a function of redshift.  We define this parameter as follows:

\begin{equation}
f_{c}=(n_o-n_t)/n_t
\label{eqn:fbg}
\end{equation}

\noindent The blue triangles and red squares in Figure \ref{fig:bgsub} indicate dwarfs and giants respectively.  The correction fractions are typically larger for the dwarfs than they are for the giants.  The giants correction is typically constrained between 0-20\% while the dwarfs exhibit a larger range in correction fraction.  Note that $f_c$ can become negative at high $z$ for some of the dwarf populations, which is expected due to increasingly fewer background galaxies in the selection box and increasingly larger colour-error; ultimately allowing the colour-error correction to dominate over the background subtraction.  We show only the clusters with background and area-corrected number dwarfs plus giants larger than 50 ($n_d+n_g > 50$).  Four clusters that are poorer than this limit can have $f_c$ values in excess of 50\% and we have excluded them from the rest of the analysis that follows.  The excluded clusters are marked with an asterisk in Table \ref{tab:targets}.

\subsection{GDR results} \label{gdrresults}

Figure \ref{fig:gdr_plain} shows the GDR as a function of redshift for all of the clusters in our sample (filled red symbols).  The data indicate that the GDR increases with redshift.  The uncertainties include contributions from the shot noise in both the cluster field and background populations and a term that accounts for the error introduced by the presence of large-scale structure in the background.  For each cluster we estimate the large-scale structure term by randomly placing 100 apertures onto the CFHTLS Deep fields with radii of 750 kpc and then calculating the 68\% inclusion limits for the number of background dwarfs and giants assuming a Poisson distribution.  The solid blue line in Figure \ref{fig:gdr_plain} shows the results of a best fit linear relationship between GDR and $z$ parameterized by $GDR=\alpha z + \beta$.  We use the MPFITEXY routine of Williams et al. (2010) to derive best fit values of $\alpha$ and $\beta$, which uses the MPFIT package by Markwardt (2009).  We also allow for some intrinsic scatter in the GDR ($\sigma_{int}$), which is allowed to vary in the fit.  The data are bootstrap sampled for 1000 realizations to obtain an empirical estimate of the uncertainties on the parameters.  We find $\alpha=0.88\pm0.15$, $\beta=0.44\pm0.03$.  The reduced $\chi^2$ of the fit is approximately equal to unity for all realizations, indicating that $\sigma_{int}$ is consistent with zero.  If the large-scale structure error is neglected, however, the best fit parameters are similar $\alpha=0.91\pm0.14$, $\beta=0.47\pm0.03$ but the required $\sigma_{int}=0.088\pm0.017$.  This suggests that large-scale structure is the dominant source of this scatter.  A Spearman's rank test (non-parametric) of the GDR vs $z$ correlation returns $\rho=0.54$ with a significance of greater than 5$\sigma$ over the null hypothesis.  The data and their best fit parameter values clearly favour an evolving GDR, but we reserve further discussion of this and a detailed comparison with results from the literature for \S \ref{litcomp}.  A list of the GDR measurements for each cluster, along with the number of dwarfs and giants both before and after the background$+$colour-error correction (see \S \ref{bgcolcor}) is given in Table \ref{tab:gdr}.

\begin{table*}
\caption{The GDR measurements.  The columns give the name, the redshift ($z$), the fully-corrected GDR ($GDR$), the statistical uncertainty on the GDR ($\sigma_{GDR}$), the number of giants and dwarfs after data corrections ($n_g$ \& $n_d$ respectively), the number of giants and dwarfs prior to data corrections ($n_g^{raw}$ \& $n_d^{raw}$ respectively), the maximum $m_{r'}$ limits used for defining the giant and dwarf samples ($m_{r'}^g$ \& $m_{r'}^d$ respectively), the colour of the red sequence ($(g'-r')_{rs}\pm0.01$) - see \S \ref{redseq}) and the X-ray temperature ($T_x$) as measured with ASCA (Horner 2001).  Additional temperature measurements are added from BAX and marked with a * in the last column.  The numbers of galaxies have been adjusted to account for the image area that was masked in the detection procedure.}
\begin{tabular}{| l | r | r | r | r | r | r | r | r | r | c | r | r | c |}
\hline
Name & \multicolumn{1}{|c|}{$z$} &\multicolumn{1}{|c|}{$GDR$} & \multicolumn{1}{|c|}{$\sigma_{GDR}$}  & \multicolumn{1}{|c|}{$n_g$} & \multicolumn{1}{|c|}{$n_g^{raw}$} & \multicolumn{1}{|c|}{$n_d$} & \multicolumn{1}{|c|}{$n_d^{raw}$} & \multicolumn{1}{|c|}{$m_{r'}^g$} &\multicolumn{1}{|c|}{$m_{r'}^d$} & \multicolumn{1}{|c|}{$(g'-r')_{rs}$} & \multicolumn{1}{|c|}{$T_x$} & \multicolumn{1}{|c|}{$\sigma_{T_x}$} & BAX\\
\hline
Abell119 &  0.04 &  0.47 &  0.13 &  33.2 &  35.6 &  70.9 &  81.3 &  16.1 &  17.9 &   0.70 &   5.93 &   0.17 &  \\
MKW3S &  0.05 &  0.53 &  0.21 &  22.2 &  24.2 &  42.0 &  53.8 &  16.2 &  18.0 &   0.73 &   3.41 &   0.05 &  \\
Abell780 &  0.05 &  0.49 &  0.19 &  22.3 &  24.3 &  45.7 &  56.3 &  16.5 &  18.3 &   0.72 &   3.54 &   0.06 &  \\
Abell754 &  0.05 &  0.74 &  0.17 &  56.9 &  59.3 &  76.4 &  87.3 &  16.6 &  18.3 &   0.72 &   9.94 &   0.31 &  \\
Abell85 &  0.05 &  0.65 &  0.18 &  39.1 &  41.3 &  60.1 &  71.5 &  16.6 &  18.4 &   0.73 &   5.73 &   0.14 &  \\
Abell2319 &  0.06 &  0.48 &  0.08 &  83.5 &  86.3 & 175.7 & 185.2 &  16.6 &  18.4 &   0.71 &   9.62 &   0.31 &  \\
Abell133 &  0.06 &  0.66 &  0.21 &  30.0 &  32.5 &  45.3 &  56.9 &  16.7 &  18.5 &   0.75 &   3.71 &   0.07 &  \\
Abell1991 &  0.06 &  0.73 &  0.26 &  27.4 &  29.7 &  37.7 &  49.9 &  16.8 &  18.6 &   0.77 &   5.40 &   3.00 & * \\
Abell1795 &  0.06 &  0.41 &  0.12 &  35.9 &  38.5 &  87.5 &  98.1 &  16.9 &  18.7 &   0.72 &   5.49 &   0.06 &  \\
Abell553 &  0.07 &  0.54 &  0.17 &  35.9 &  38.2 &  66.3 &  79.2 &  17.0 &  18.8 &   0.75 &   &     &  \\
Abell644 &  0.07 &  0.37 &  0.09 &  42.9 &  46.0 & 116.5 & 129.1 &  17.2 &  19.0 &   0.78 &   7.31 &   0.14 &  \\
Abell399 &  0.07 &  0.42 &  0.10 &  48.3 &  51.5 & 115.2 & 125.5 &  17.2 &  19.0 &   0.71 &   6.99 &   0.23 &  \\
Abell2065 &  0.07 &  0.74 &  0.14 &  82.0 &  85.4 & 110.6 & 122.4 &  17.2 &  19.0 &   0.77 &   5.35 &   0.12 &  \\
Abell401 &  0.07 &  0.54 &  0.14 &  51.3 &  54.3 &  94.6 & 107.4 &  17.3 &  19.1 &   0.75 &   8.07 &   0.20 &  \\
ZwCl1215 &  0.08 &  0.44 &  0.13 &  33.9 &  37.1 &  76.8 &  88.6 &  17.3 &  19.1 &   0.80 &   6.54 &   0.21 & * \\
Abell2670 &  0.08 &  0.91 &  0.22 &  57.7 &  61.0 &  63.1 &  76.0 &  17.3 &  19.1 &   0.78 &   3.98 &   0.18 &  \\
Abell2029 &  0.08 &  0.54 &  0.12 &  52.2 &  55.5 &  97.3 & 110.0 &  17.4 &  19.2 &   0.78 &   7.30 &   0.37 &  \\
Abell2495 &  0.08 &  0.41 &  0.13 &  29.7 &  33.2 &  72.4 &  84.0 &  17.4 &  19.2 &   0.77 &     &     &  \\
RXSJ2344-04 &  0.08 &  0.36 &  0.15 &  18.5 &  21.5 &  51.4 &  64.6 &  17.4 &  19.2 &   0.78 &     &     &  \\
ZwCl0628 &  0.08 &  0.42 &  0.10 &  54.5 &  58.0 & 128.6 & 140.5 &  17.5 &  19.3 &   0.73 &  6.20 &   2.50 & * \\
Abell2033 &  0.08 &  0.56 &  0.17 &  33.1 &  36.5 &  59.5 &  73.1 &  17.5 &  19.3 &   0.76 &   4.16 &   0.12 &  \\
Abell1650 &  0.08 &  0.34 &  0.11 &  30.0 &  33.7 &  88.1 & 100.0 &  17.6 &  19.3 &   0.76 &   5.89 &   0.12 &  \\
Abell1651 &  0.08 &  0.51 &  0.14 &  37.7 &  41.0 &  74.3 &  87.0 &  17.6 &  19.4 &   0.76 &   5.87 &   0.15 &  \\
Abell2420 &  0.09 &  0.68 &  0.18 &  47.9 &  51.2 &  70.1 &  84.6 &  17.6 &  19.4 &   0.78 &   6.00 &   1.60 & * \\
Abell2597 &  0.09 &  0.72 &  0.27 &  24.0 &  27.0 &  33.1 &  46.0 &  17.6 &  19.4 &   0.81 &   3.58 &   0.07 &  \\
Abell478 &  0.09 &  0.99 &  0.22 &  72.1 &  74.3 &  73.2 &  83.6 &  17.7 &  19.5 &   0.67 &   7.07 &   0.17 &  \\
Abell2440 &  0.09 &  0.57 &  0.15 &  41.8 &  44.9 &  72.7 &  85.6 &  17.7 &  19.5 &   0.82 &   4.31 &   0.15 &  \\
Abell2142 &  0.09 &  0.57 &  0.14 &  54.5 &  57.1 &  95.8 & 109.1 &  17.8 &  19.5 &   0.82 &   8.24 &   0.30 &  \\
Abell1927 &  0.09 &  0.41 &  0.14 &  25.9 &  29.0 &  62.6 &  77.0 &  17.9 &  19.7 &   0.82 &     &     &  \\
Abell21 &  0.09 &  0.58 &  0.15 &  46.1 &  48.9 &  79.2 &  91.6 &  17.9 &  19.7 &   0.80 &     &     &  \\
Abell2426 &  0.10 &  0.55 &  0.13 &  54.0 &  58.0 &  97.4 & 110.0 &  17.9 &  19.7 &   0.78 &     &     &  \\
Abell2055 &  0.10 &  0.44 &  0.14 &  28.5 &  32.0 &  64.8 &  78.0 &  18.0 &  19.8 &   0.76 &   5.80 &    & * \\
A98 &  0.10 &  0.64 &  0.18 &  41.0 &  44.5 &  64.1 &  77.4 &  18.1 &  19.8 &   0.75 &     &     &  \\
Abell1285 &  0.11 &  0.48 &  0.13 &  45.0 &  48.7 &  94.1 & 108.5 &  18.1 &  19.9 &   0.82 &   4.10 &   3.00 & * \\
Abell7 &  0.11 &  0.63 &  0.18 &  38.9 &  42.3 &  61.3 &  74.5 &  18.1 &  19.9 &   0.77 &     &     &  \\
Abell2443 &  0.11 &  0.44 &  0.13 &  37.7 &  41.3 &  86.4 &  98.2 &  18.2 &  19.9 &   0.81 &     &     &  \\
RXCJ0352+19 &  0.11 &  0.44 &  0.20 &  19.4 &  23.1 &  44.5 &  56.6 &  18.2 &  20.0 &   0.76 &     &     &  \\
Abell2703 &  0.11 &  0.38 &  0.15 &  26.1 &  29.9 &  68.2 &  83.9 &  18.3 &  20.1 &   0.84 &     &     &  \\
Abell2069 &  0.12 &  0.54 &  0.13 &  47.8 &  52.0 &  88.0 & 103.0 &  18.3 &  20.1 &   0.83 &   6.30 &   0.20 & * \\
Abell1361 &  0.12 &  0.42 &  0.18 &  19.1 &  22.6 &  45.9 &  60.7 &  18.3 &  20.1 &   0.82 &     &     &  \\
Abell2050 &  0.12 &  0.71 &  0.18 &  51.3 &  55.0 &  72.4 &  85.3 &  18.4 &  20.1 &   0.87 &   4.34 &   0.07 & * \\
RXCJ0736+39 &  0.12 &  0.51 &  0.22 &  19.5 &  24.1 &  38.2 &  52.4 &  18.4 &  20.1 &   0.83 &     &     &  \\
Abell1348 &  0.12 &  0.46 &  0.16 &  25.7 &  29.5 &  55.5 &  71.3 &  18.4 &  20.2 &   0.84 &   3.60 &   0.08 & * \\
Abell961 &  0.12 &  0.44 &  0.14 &  30.1 &  33.6 &  67.8 &  83.6 &  18.5 &  20.3 &   0.85 &     &     &  \\
Abell1033 &  0.13 &  0.51 &  0.13 &  45.3 &  48.9 &  89.7 & 104.8 &  18.5 &  20.3 &   0.81 &     &     &  \\
Abell2627 &  0.13 &  0.37 &  0.16 &  17.8 &  22.0 &  48.1 &  65.0 &  18.5 &  20.3 &   0.81 &     &     &  \\
Abell655 &  0.13 &  0.57 &  0.13 &  66.4 &  70.8 & 116.6 & 133.1 &  18.5 &  20.3 &   0.84 &     &     &  \\
\hline
\end{tabular}
\label{tab:gdr}
\end{table*}
\begin{table*}
\caption{Table \ref{tab:gdr} continued...}
\centering
\begin{tabular}{| l | r | r | r | r | r | r | r | r | r | c | r | r | c |}
\hline
Name & \multicolumn{1}{|c|}{$z$} &\multicolumn{1}{|c|}{$GDR$} & \multicolumn{1}{|c|}{$\sigma_{GDR}$}  & \multicolumn{1}{|c|}{$n_g$} & \multicolumn{1}{|c|}{$n_g^{raw}$} & \multicolumn{1}{|c|}{$n_d$} & \multicolumn{1}{|c|}{$n_d^{raw}$} & \multicolumn{1}{|c|}{$m_{r'}^g$} &\multicolumn{1}{|c|}{$m_{r'}^d$} & \multicolumn{1}{|c|}{$(g'-r')_{rs}$} & \multicolumn{1}{|c|}{$T_x$} & \multicolumn{1}{|c|}{$\sigma_{T_x}$} & BAX\\
\hline
Abell646 &  0.13 &  0.69 &  0.26 &  23.7 &  27.3 &  34.3 &  51.5 &  18.6 &  20.4 &   0.85 &     &     &  \\
Abell1132 &  0.14 &  0.44 &  0.11 &  45.3 &  49.3 & 103.2 & 119.6 &  18.7 &  20.5 &   0.85 &     &     &  \\
Abell795 &  0.14 &  0.67 &  0.15 &  66.3 &  70.2 &  99.5 & 115.1 &  18.7 &  20.5 &   0.87 &     &     &  \\
A1882 &  0.14 &  0.60 &  0.18 &  38.2 &  42.3 &  63.5 &  79.9 &  18.7 &  20.5 &   0.85 &     &     &  \\
Abell1068 &  0.14 &  0.43 &  0.15 &  26.5 &  31.0 &  61.0 &  78.0 &  18.7 &  20.5 &   0.85 &   3.87 &   0.12 &  \\
Abell1413 &  0.14 &  0.75 &  0.16 &  62.7 &  67.0 &  84.0 & 101.0 &  18.8 &  20.6 &   0.86 &   7.09 &   0.25 &  \\
ZwCl1023 &  0.14 &  0.62 &  0.21 &  29.2 &  33.0 &  47.2 &  63.0 &  18.8 &  20.6 &   0.89 &     &     &  \\
Abell990 &  0.14 &  0.65 &  0.19 &  45.1 &  48.9 &  69.1 &  87.4 &  18.8 &  20.6 &   0.87 &   5.75 &   0.22 &  \\
Abell2409 &  0.15 &  0.90 &  0.20 &  67.4 &  71.0 &  75.2 &  91.0 &  18.9 &  20.7 &   0.87 &   5.50 &   0.25 & * \\
A2204 &  0.15 &  0.79 &  0.17 &  91.8 &  95.8 & 116.7 & 134.8 &  18.9 &  20.7 &   0.86 &   7.41 &   0.26 &  \\
A545 &  0.15 &  0.84 &  0.15 &  89.4 &  93.5 & 107.0 & 124.7 &  19.0 &  20.8 &   0.86 &   5.50 &   3.00 & * \\
A2104 &  0.16 &  0.65 &  0.13 &  69.5 &  74.1 & 106.2 & 126.2 &  19.1 &  20.9 &   0.85 &   9.31 &   0.47 &  \\
A2259 &  0.16 &  0.57 &  0.14 &  44.0 &  48.0 &  77.1 &  96.1 &  19.1 &  20.9 &   0.90 &   5.32 &   0.27 &  \\
A1234 &  0.17 &  0.58 &  0.16 &  39.6 &  43.7 &  68.5 &  87.4 &  19.1 &  20.9 &   0.87 &     &     &  \\
A1914 &  0.17 &  0.65 &  0.13 &  71.9 &  75.7 & 109.7 & 128.1 &  19.3 &  21.0 &   0.99 &   9.48 &   0.45 &  \\
A586 &  0.17 &  0.66 &  0.15 &  71.0 &  75.2 & 107.5 & 126.8 &  19.3 &  21.0 &   0.97 &   6.39 &   0.60 &  \\
A1246 &  0.19 &  0.80 &  0.16 &  76.4 &  81.7 &  95.4 & 117.1 &  19.5 &  21.3 &   0.94 &   6.04 &   0.37 &  \\
MS0440 &  0.19 &  0.58 &  0.18 &  43.6 &  48.2 &  75.6 &  96.5 &  19.5 &  21.3 &   0.94 &   5.02 &   0.50 &  \\
A115 &  0.20 &  0.79 &  0.19 &  55.8 &  60.9 &  71.1 &  92.5 &  19.7 &  21.4 &   1.00 &   6.45 &   0.31 &  \\
A115N &  0.20 &  0.52 &  0.15 &  45.8 &  50.4 &  88.0 & 108.7 &  19.7 &  21.4 &   0.98 &   6.45 &   0.31 &  \\
A2163 &  0.20 &  0.71 &  0.10 & 128.7 & 133.6 & 182.5 & 204.6 &  19.6 &  21.4 &   0.88 &  12.12 &   0.57 &  \\
A2261 &  0.20 &  0.54 &  0.11 &  67.2 &  71.0 & 124.8 & 142.0 &  19.7 &  21.5 &   1.10 &   6.88 &   0.41 &  \\
A520 &  0.20 &  0.68 &  0.15 &  57.7 &  62.1 &  85.0 & 105.1 &  19.7 &  21.4 &   1.02 &   7.81 &   0.64 &  \\
A223a &  0.21 &  0.71 &  0.17 &  49.7 &  54.0 &  70.5 &  89.0 &  19.8 &  21.5 &   1.01 &   5.12 &   0.66 &  \\
A223b &  0.21 &  0.51 &  0.14 &  38.7 &  43.0 &  75.1 &  94.0 &  19.8 &  21.5 &   1.03 &   5.12 &   0.66 &  \\
A1942 &  0.22 &  0.62 &  0.15 &  51.2 &  55.5 &  82.3 & 100.8 &  19.9 &  21.6 &   1.05 &   5.12 &   0.56 &  \\
A2111 &  0.23 &  0.65 &  0.14 &  70.4 &  75.0 & 107.7 & 126.3 &  20.0 &  21.8 &   1.07 &   8.02 &   0.77 &  \\
A2125 &  0.25 &  0.94 &  0.22 &  61.4 &  64.5 &  65.3 &  78.3 &  20.3 &  22.1 &   1.36 &   2.60 &   0.10 & * \\
A1835 &  0.25 &  0.68 &  0.13 &  75.9 &  80.2 & 111.8 & 126.2 &  20.2 &  22.0 &   1.17 &   7.65 &   0.31 &  \\
A521 &  0.25 &  0.71 &  0.16 &  62.1 &  67.0 &  87.7 & 104.0 &  20.2 &  21.9 &   1.14 &   6.74 &   0.45 &  \\
CL1938 &  0.26 &  0.75 &  0.14 &  89.0 &  93.2 & 118.9 & 132.0 &  20.4 &  22.1 &   1.22 &   7.52 &   0.37 &  \\
A1758b &  0.28 &  0.52 &  0.11 &  64.4 &  69.0 & 123.3 & 137.0 &  20.5 &  22.2 &   1.23 &   7.95 &   0.62 &  \\
A697 &  0.28 &  0.57 &  0.12 &  68.4 &  72.9 & 120.7 & 133.4 &  20.5 &  22.2 &   1.28 &   9.14 &   0.54 &  \\
A611 &  0.29 &  0.74 &  0.17 &  57.6 &  62.4 &  78.3 &  91.6 &  20.6 &  22.3 &   1.24 &   6.69 &   0.44 &  \\
A959 &  0.29 &  0.84 &  0.17 &  73.4 &  78.0 &  87.7 & 101.0 &  20.5 &  22.3 &   1.19 &   6.26 &   0.71 &  \\
A2537 &  0.30 &  0.83 &  0.15 &  97.2 & 102.0 & 116.8 & 129.0 &  20.7 &  22.5 &   1.32 &   6.08 &   0.49 &  \\
MS1008 &  0.30 &  0.92 &  0.21 &  66.4 &  71.5 &  72.1 &  84.6 &  20.7 &  22.5 &   1.29 &   7.47 &   1.21 &  \\
A370 &  0.38 &  0.93 &  0.17 &  91.6 &  93.0 &  98.6 & 100.0 &  21.4 &  23.1 &   1.54 &   7.20 &   0.77 &  \\
A851 &  0.41 &  1.14 &  0.26 &  72.8 &  77.0 &  63.8 &  73.0 &  21.6 &  23.4 &   1.47 &   7.21 &   1.34 &  \\
RXJ0856 &  0.41 &  0.82 &  0.19 &  69.2 &  72.0 &  84.3 &  84.0 &  21.7 &  23.4 &   1.47 &     &     &  \\
RXJ2228 &  0.41 &  0.87 &  0.19 &  73.8 &  76.0 &  85.2 &  85.0 &  21.6 &  23.3 &   1.51 &   7.90 &   0.60 & * \\
RXJ1003 &  0.42 &  0.76 &  0.21 &  47.0 &  48.0 &  61.9 &  60.0 &  21.7 &  23.5 &   1.57 &     &     &  \\
MS1621 &  0.43 &  0.61 &  0.14 &  63.1 &  65.0 & 102.8 &  99.0 &  21.8 &  23.5 &   1.53 &   6.54 &   1.02 &  \\
CL0910 &  0.44 &  0.64 &  0.18 &  43.4 &  47.5 &  67.8 &  76.2 &  21.9 &  23.6 &   1.51 &   6.61 &   0.60 &  \\
MACS1206 &  0.44 &  0.69 &  0.15 &  78.0 &  81.0 & 113.2 & 103.0 &  21.9 &  23.6 &   1.50 &  10.20 &   1.00 & * \\
RXJ1347 &  0.45 &  0.67 &  0.15 &  57.9 &  62.0 &  87.0 &  95.0 &  22.0 &  23.8 &   1.57 &  10.88 &   0.66 &  \\
RXJ1701 &  0.45 &  1.05 &  0.35 &  42.9 &  45.2 &  40.8 &  43.2 &  22.0 &  23.7 &   1.54 &   4.50 &   1.00 & * \\
3C295 &  0.46 &  1.01 &  0.31 &  41.1 &  45.0 &  40.8 &  48.0 &  22.0 &  23.6 &   1.48 &   6.51 &   0.99 &  \\
RXJ1524 &  0.52 &  0.89 &  0.28 &  47.4 &  50.3 &  53.5 &  58.0 &  22.0 &  23.4 &   1.39 &   5.10 &   0.36 &  \\
MS0451 &  0.54 &  1.43 &  0.33 &  86.7 &  93.0 &  60.5 &  67.0 &  22.7 &  24.0 &   1.46 &   8.62 &   1.21 &  \\
\hline
\end{tabular}
\label{tab:gdr2}
\end{table*}

\subsection{Sensitivity to Analysis Parameters} \label{selcut}

One of the issues that complicates the comparison of GDR (and dwarf-to-giant ratio) between studies is the varying selection criteria used to define the dwarf and/or giant populations.  The $K$-correction method, the projected cluster-centric distance within which to include galaxies in the dwarf and giant populations as well as the maximum colour-residual with respect to the CMR that defines a red sequence member are examples of parameters that often vary between different studies.  In this section, we compute the GDR using a range of selection criteria to see how much of the apparent discrepancy can be explained through these choices.

\begin{figure*}
\begin{flushleft}
\begin{minipage}[b]{0.28\linewidth}
    \includegraphics[width=2.6in]{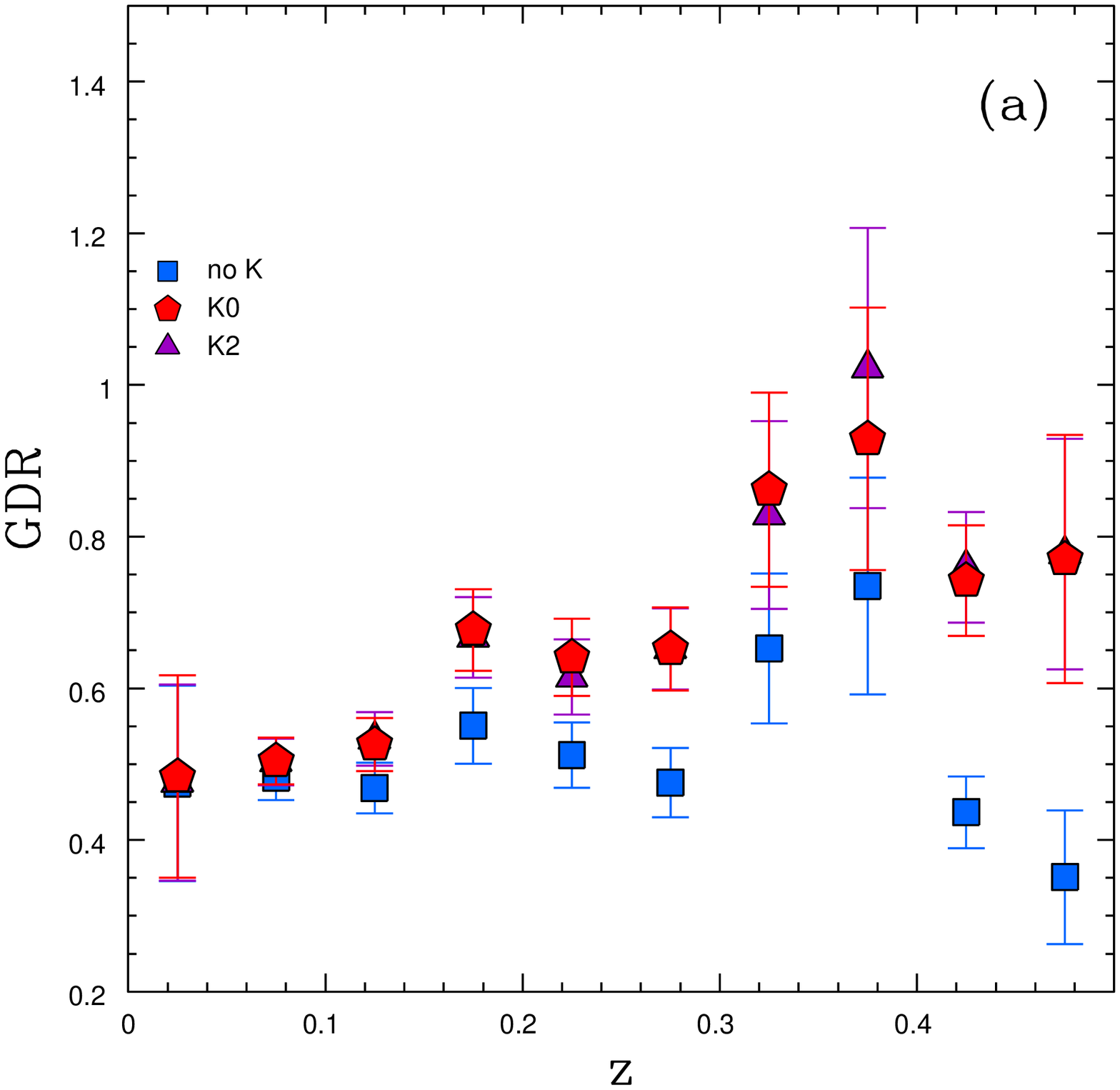}
\end{minipage}
\hspace{0.5cm}
\begin{minipage}[b]{0.28\linewidth}
    \includegraphics[width=2.6in]{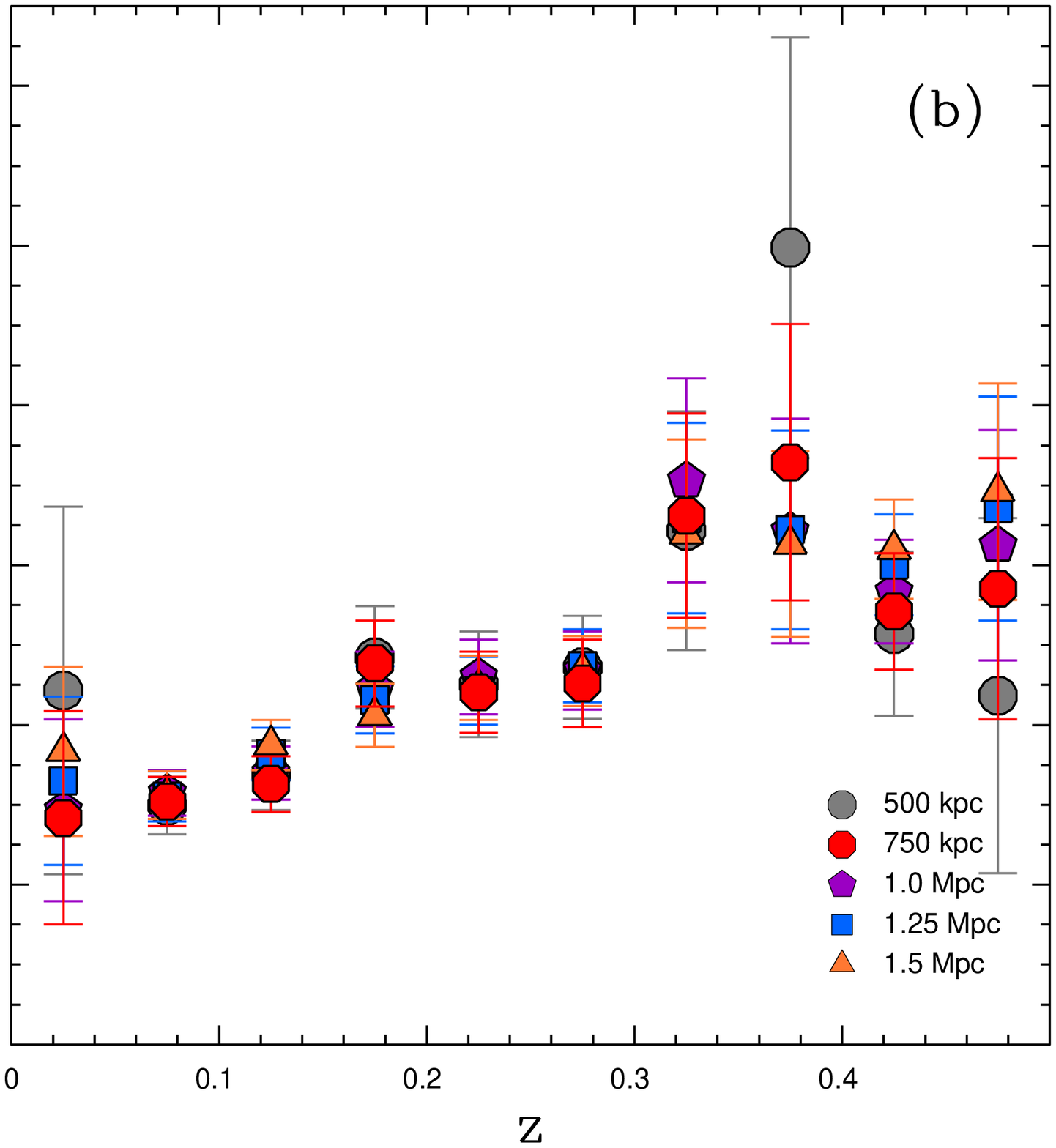}
\end{minipage}
\hspace{0.5cm}
\begin{minipage}[b]{0.28\linewidth}
    \includegraphics[width=2.6in]{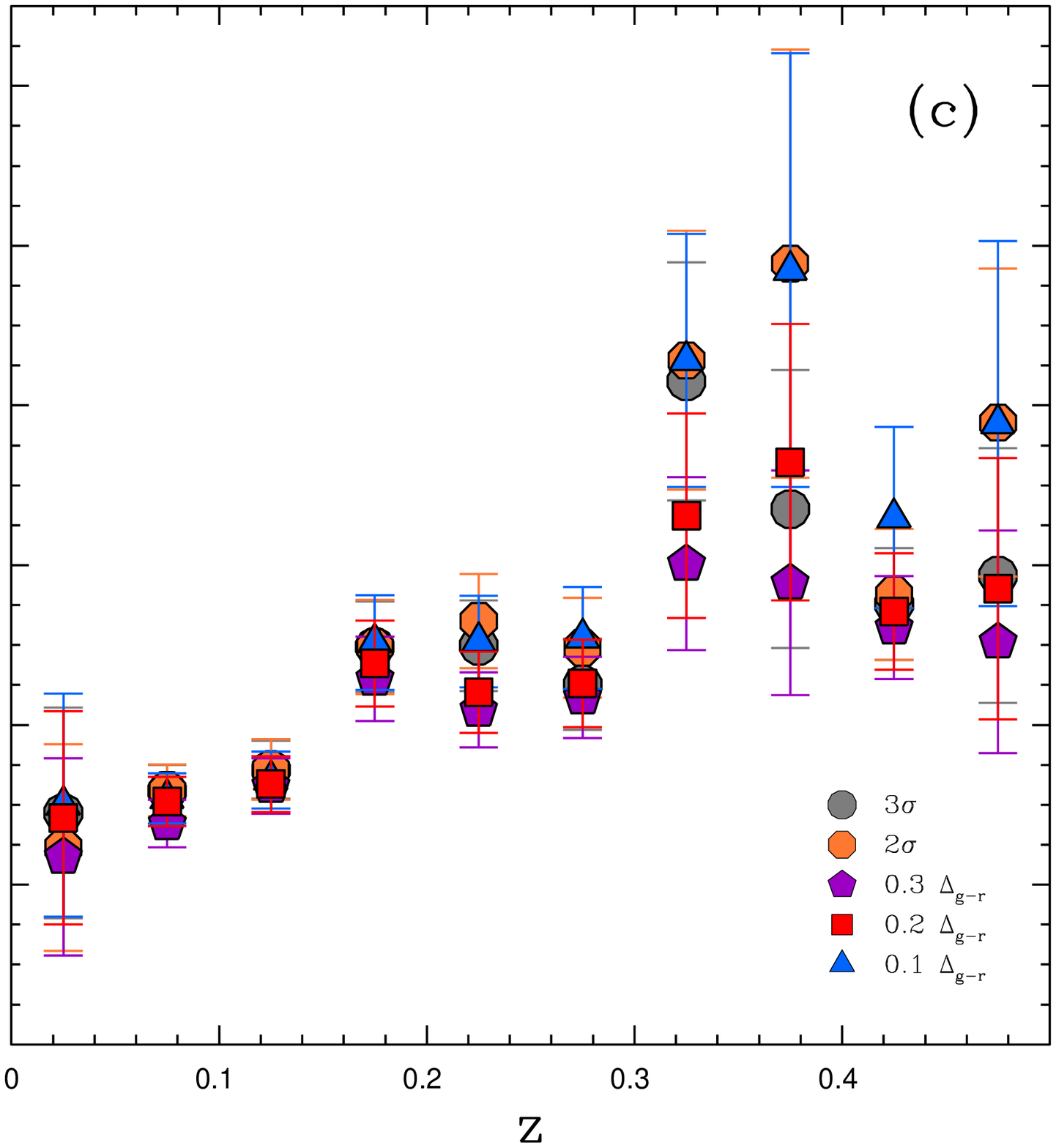}
\end{minipage}

\caption{\textbf{(a)} The giant-to-dwarf ratio using various methods of $K$-correction.  The different symbols correspond to 2 different $K$-correction methods (see text for details) as well the effect of neglecting the $K$-correction.  \textbf{(b)} The giant-to-dwarf ratio using various apertures sizes for red sequence membership.  The different symbols refer to galaxies within $r_{lim}$ in projected distance from the BCG with $r_{lim}=$500 kpc, 750 kpc, 1000 kpc, 1250 kpc and 1500 kpc.  \textbf{(c)} The giant-to-dwarf ratio as computed using various colour selection criteria.  The different symbols show $|\Delta(g'-r')|<0.1$, $|\Delta(g'-r')|<0.2$, $|\Delta(g'-r')|<0.3$, $|\Delta(g'-r')|/\sigma<2$ and $|\Delta(g'-r')|/\sigma<3$.}
\label{fig:gdr_selcut}
\end{flushleft}
\end{figure*}

To examine how our results depend on the adopted $K$-correction described above, we compare the GDRs obtained using two different $K$-correction methods.  The results are plotted in Figure \ref{fig:gdr_selcut}a.  The first $K$-correction method (K0) is the one outlined above using the CB07 models and assumes solar metallicity and a template age that is determined by matching the observed $(g'-r')$ colour.  The next method (K2) uses the $K$-correction code of Chilingarian, Melchior and Zolotukhin (2010) which has been shown to agree with the empirically derived corrections of Blanton \& Roweis (2007) up to z=0.5.  We also show the results of neglecting the $K$-correction completely (no K).

The differences in GDR between one $K$-correction method and another are typically smaller than the GDR error in an individual redshift bin.  Moreover, the selection of either K-correction method does not affect the conclusions regarding the presence or absence of a trend with redshift, only how strong this evolution might be.  However, neglecting the $K$-correction would lead to a contradictory view of the redshift evolution in GDR.

Although it is not shown in Figure \ref{fig:gdr_selcut}a, we also test the effect of neglecting the evolution correction (ie. $E=0$, see \S \ref{kpluse}).  We find that contrary to neglecting the $K$-correction, this leads to a significant steepening of the GDR vs $z$ relation ($\alpha = 1.61 \pm 0.15$).  Therefore, to obtain a flat GDR vs $z$ by varying the $E$-correction would require $E$ to be parameterized with a significantly stronger redshift dependence than what we have assumed here.  Such corrections are unreasonable for the early-type galaxies that make up the red sequence.

Figure \ref{fig:gdr_selcut}b shows the effect on GDR of varying the aperture size ($r_{lim}$), centred on the BCG, within which to count red sequence galaxies.  The data shown in the figure are for galaxies within a projected radius of 500 kpc, 750 kpc, 1000 kpc, 1250 kpc and 1500 kpc.  The variation in the results, most notable in the lowest redshift bin, is all within the error in each individual bin.  This shows that the GDR is insensitive to aperture choice on these scales (up to a factor of 3 in BCG-centric distance).  Choosing instead a scaled radius (eg. $r_{500}$) to measure GDR would yield similar results.  Using $r_{500}=2.48 (T_x/10 \textrm{ keV})^{1/2}$ (Evrard et al. 1996) and the $T_x$ data shown in table \ref{tab:gdr}, we calculate that the mean value of $r_{500}$ changes by at most a factor of 1.33 when comparing any pair of redshift bins.  Since this is well within the factor of 3 range in aperture size over which the GDR measurement is robust, we conclude that the use of scaled radii to measure GDR does not alter the conclusions regarding GDR evolution.  We also note that the largest apertures show the smallest GDR errors.  This is expected since they are based on Poisson statistics and the GDR is a ratio.  As the number of galaxies in the population increases, the error as a percentage on that number decreases.  We argue, however, that it is better to use a smaller aperture such as $r_{lim}=750$kpc because the potential for systematic error introduced by the background subtraction procedure, which is not included in the error estimates shown in Figure \ref{fig:gdr_selcut}b, grows as the aperture size increases.

It is also interesting to examine the differences in the GDR between the cluster centre and cluster outskirts.  Such a discussion does not fit well in this section however, which is intended to assess the possible variation in GDR introduced by commonly used selection criteria.  For a discussion of the GDR in the inner vs outer parts of clusters we refer the reader to \S \ref{driver}.

We consider the possibility that BCG position may be a poor indicator of the cluster centre as a function of redshift.  Depending on the radial profile of the GDR, this effect could potentially mimic GDR evolution.  To test this, we measure the GDR in $0.05 < z < 0.15$ clusters using a centre position that is offset between 0 to 6.5 arcminutes from the BCG.  This shift corresponds roughly to 750 kpc at $z=0.1$.  The results are consistent with those obtained using the BCG as the cluster centre, indicating that the GDR evolution is unaffected by centreing errors on these angular scales.

We show in Figure \ref{fig:gdr_selcut}c how varying the colour selection criteria affects the measurement of the GDR.  Here we compare several cases of two distinct selection criteria.  The first requires galaxies to be within $|\Delta(g'-r')|<0.2$, which is the assumed criteria throughout the rest of this work, but we also show here the results for $|\Delta(g'-r')|<0.1$ and $|\Delta(g'-r')|<0.3$.  The second selection method that we investigate is based on the observed scatter in $(g'-r')$ of the red sequence ($\sigma$), assuming a gaussian distribution about the best fit colour-magnitude relation.  Figure \ref{fig:gdr_selcut}c shows the GDR of galaxies that lie within $|\Delta(g'-r')|/\sigma<3$ and $|\Delta(g'-r')|/\sigma<2$.  The effect of colour selection criteria is larger than that introduced by the $r_{lim}$ selection criteria and comparable to the effect of varying the $K$-correction method.  However, the change in GDR is always smaller than the error in an individual redshift bin, which demonstrates that the choice of a particular colour criterion does not affect conclusions about the presence of evolution.

\subsection{Comparison with Literature} \label{litcomp}

 \begin{figure}
    \centering
    \includegraphics[width=3.2in]{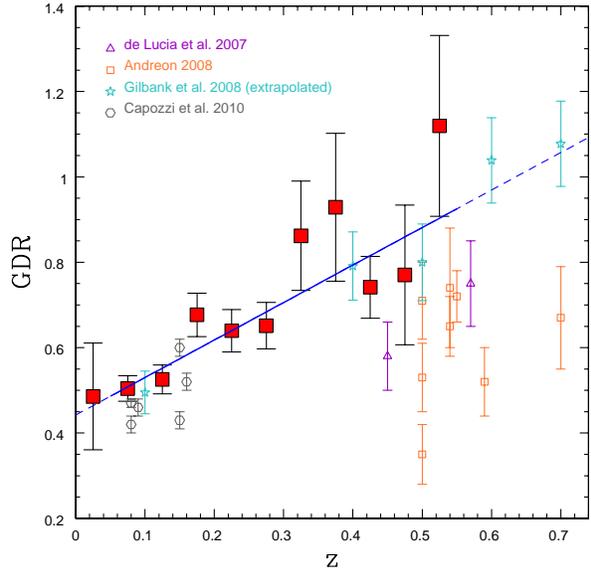}
    \caption{The ratio of the number density of giant galaxies to that of dwarf galaxies as a function of redshift.  The data has been binned in redshift bins of $\Delta z = 0.05$ (red squares).  We also show various data from the literature.  Note that the Gilbank et al. (2008) values have been adjusted to be comparable directly with ours (see text for description).}
    \label{fig:gdr}
 \end{figure}

The disagreements in the literature regarding the evolution of the GDR/DGR are primarily driven by the interpretation of various results and not necessarily by disagreements among the data themselves.  In a study based on DGR data culled from the available literature, Gilbank and Balogh (2008) show that if one neglects the $z>1$ clusters of Andreon (2008), where the filters do not bracket the 4000 angstrom break, DGR vs $z$ data is fit reasonably well by a single power law in $(1+z)$.  Andreon (2008) puts forth an important caution that one must be careful to account for the presence of an intrinsic scatter in DGR from one cluster to another so as not to be biased by outliers.  Andreon (2008) reports an intrinsic scatter $\sigma_{int}=0.13 \pm 0.06$, arguing that this is enough to reject trends as steep $\alpha \sim 1.3$ found in de Lucia et al. (2007), where $\sigma_{int}$ was not taken into account. The author also provides a caveat that the observed spiral morphologies of some nearby red sequence galaxies (eg. Butcher \& Oemler 1984) indicate recent star formation, colour transformation and red sequence buildup.  However, the associated GDR evolution, if any, is not steep enough to be detected in small samples of $\sim 25$ clusters.  To contribute to this discussion we compare these result with our own for $\sim 100$ clusters.

As discussed in \S \ref{gdrresults} we find $\alpha=0.88\pm0.15$, $\beta=0.44\pm0.03$ and $\sigma_{int}$ consistent with zero.  If the large scale structure term is neglected in the GDR uncertainty we obtain a $\sigma_{int}=0.088\pm0.017$ which is consistent with the value inferred by Andreon (2008).  Our best fit slope is significant, though not as steep as that found in de Lucia et al. (2007).  Figure \ref{fig:gdr} shows how our results compare with those from the literature.  The large square symbols indicate the error-weighted mean GDR for our sample in several redshift bins.  The solid blue line shows our best fit linear relation described above.  Data taken from the literature are overplotted.  The Gilbank et al. (2008) results are based on slightly different magnitude limits for defining dwarfs and giants than those used here.  To make a fairer comparison we compute the GDR by applying their magnitude cuts to our data, then calculate a correction based on the difference between the best fit $\alpha$ and $\beta$ from those obtained using our default cuts.  We apply this correction to the Gilbank et al. (2008) values and plot the `extrapolated' values in Figure \ref{fig:gdr}.  There is very good agreement between our results and the `extrapolated' values of Gilbank et al. (2008) despite the difference in redshift range and cluster masses probed.  Our GDR values are also in agreement with the low redshift GDR results of Capozzi et al. (2010), which are based on SDSS cluster catalogs and the de Lucia et al. (2007) magnitude limits.  When compared directly to the de Lucia et al. (2007) results our data do not seem to agree.  However, there is an important difference between the two analyses.  The magnitudes used in this paper are taken from SExtractor's MAG\_AUTO parameter, which attempts to follow the shape of the galaxy surface brightness profiles.  The de Lucia et al. (2007) magnitudes are extracted using apertures of 1.0", significantly smaller than the sizes within which our magnitudes are calculated.  Using a 1.0" fixed aperture truncates the galaxies at smaller radii, effectively reducing their fluxes.  This imparts a stronger bias against the giants than dwarfs, since the former are intrinsically more extended on the sky leading to a reduction in the value of the GDR and a steepening of the GDR vs $z$ relation.  We are not able to fully reproduce an analogous GDR measurement to that of de Lucia et al. (2007) because of the different angular resolution scales of CFHT and HST.  As an instructional exercise we remeasure our GDR values using a 3.0" aperture size for galaxy magnitudes, which is more suitable given the spatial resolution of CFHT.  We verify that this leads to a significant reduction in the GDR values compared to our default extraction method.  The resulting best fit to the fixed-aperture GDR vs $z$ relation has a slope $\alpha=1.90$, which represents a steepening of $\Delta \alpha \sim 1$.  We also note that because this effect operates more noticeably at lower redshifts as the angular sizes of galaxies grow and the truncation becomes more extreme, it may provide an explanation for the very steep redshift trend seen by de Lucia et al. (2007).

  \begin{figure}
    \centering
    \includegraphics[width=3.2in]{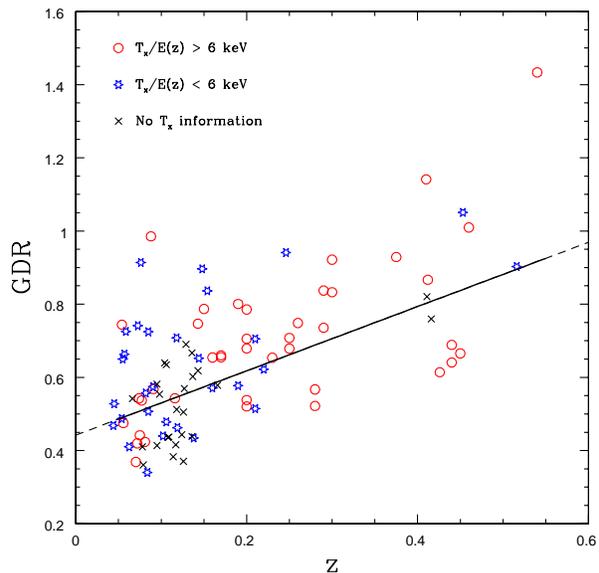}
    \caption{GDR versus $z$ with clusters coded by X-ray temperature information.  The solid black line shows the best fit linear relation found in \S \ref{gdrresults}.  The GDR evolution can be represented by a single function of redshift regardless of $T_x/E(z)$ for the full range of temperature and redshift probed by our sample.}
    \label{fig:gdr_t3}
 \end{figure}

\section{Dependence on Cluster Mass} \label{clusterprops}

As we discuss in the previous section, the GDR data can be modelled by a linear trend with $z$.  After accounting for large-scale structure in the background, we find no intrinsic scatter.  This is qualitatively consistent with the results of Lu et al. (2009) who find that the large-scale structure uncertainty comprises a significant portion of the uncertainty in DGR for individual clusters.  This result is consistent with a GDR that is a function of redshift only and is independent of any other global cluster properties, at least over the range of redshift and cluster mass investigated in this paper.  Several authors in the literature, however, have presented evidence for a dependance of GDR on global properties related to cluster mass.  For instance, de Lucia et al. (2007) find a higher GDR in their $\sigma > 600$ km/s sample than that in their $\sigma < 600$ km/s sample, with a difference in GDR of about 0.1 to 0.2.  They also analyze data from the SDSS for clusters at $z\sim0$ and find that this $\sigma$ dependence is reversed.  Gilbank et al. (2008) find an elevated and flatter GDR for poorer clusters.  Capozzi et al. (2010) argue that there may be a weak correlation between the GDR and cluster X-ray luminosity.  Furthermore, as the GDR is different in the field (Gilbank \& Balogh 2008), some mass dependence might be expected.  Because our cluster sample is not a representative sample, it may be that some portion of the observed GDR evolution arises as a consequence of potential covariance between redshift and cluster mass (more high mass systems at high $z$).  However, given the good agreement between our GDR results and those based on cluster samples with lower mean mass (eg. Gilbank et al. 2008 and Capozzi et al. 2010) we do not expect this to be the case.  Nonetheless, in this section we investigate this possibility in more detail.

  \begin{figure}
    \centering
    \includegraphics[width=3.2in]{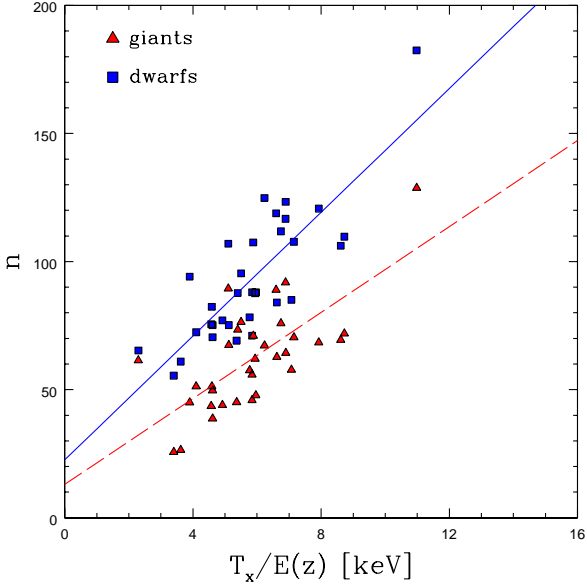}
    \caption{The number of giants (red triangles) and dwarfs (blue squares) as a function of cluster $T_x/E(z)$ for clusters in the range $0.1 < z < 0.3$.  Solid and dashed lines correspond to the best fit relations for dwarfs and giants respectively.}
    \label{fig:ngd_tx}
 \end{figure}

Many of the previous results that indicate a dependence on cluster mass are based on using cluster richness as a mass proxy.  One problem with using richness is that there is a large scatter in the mass-richness relation of up to $\pm 150$\% in mass at a given richness (Rozo et al. 2009).  Moreover, the richness, which is approximately the number of dwarfs plus giants ($n_d+n_g$) and the GDR ($n_g/n_d$) are by construction naturally covariant.  A better proxy for cluster mass is the X-ray temperature of the hot intra-cluster medium ($T_x$), which predicts the mass with significantly lower scatter (36\% in mass at a fixed $T_x$ inside $r_{500}$) and is independent of galaxy counting.

A subset of the clusters in our sample have been observed with the Advanced Satellite for Cosmology and Astrophysics (ASCA) and analyzed by Horner (2001).  For maximum consistency we take $T_x$ values from the Horner (2001) catalog for as many of our clusters as possible (59 systems).  We supplement this with an additional 15 $T_x$ measurements from the BAX X-ray galaxy cluster database\footnote{http://bax.ast.obs-mip.fr/} giving us a total of 72.  A complete listing of the $T_x$ data is given in Table \ref{tab:gdr}.  In the analysis that follows, we scale the $T_x$ values by $E(z)=\sqrt{\Omega_m(1+z)^3+\Lambda}$  to account for evolution in the cosmic background density.  We make no attempt to adjust the X-ray data for the presence of cool-cores or cluster-cluster mergers.

To investigate the effect of $T_x/E(z)$ on GDR evolution we show in Figure \ref{fig:gdr_t3} the plot of GDR vs $z$ after separating the data into 3 different $T_x/E(z)$ categories.  The data points in Figure \ref{fig:gdr_t3} coded by $T_x/E(z) > 6 keV$ (red circles), $T_x/E(z) < 6 keV$ (blue snowflakes) and no $T_x$ information (black crosses).  Error bars have been omitted for clarity but are listed in Table \ref{tab:gdr}.  Figure \ref{fig:gdr_t3} shows that indeed there are more high $T_x/E(z)$ systems at the high $z$ end of our sample and conversely more low $T_x/E(z)$ systems at the low $z$ end.  Despite this selection effect however, the data in each $T_x/E(z)$ category are consistent with the same best fit GDR vs $z$ relation found in section \ref{gdrresults}.  Specifically, we find that the best fit parameters are ($\alpha=0.92\pm0.26$, $\beta=0.49\pm0.06$) for the high $T_x/E(z)$ clusters, ($\alpha=0.90\pm0.34$, $\beta=0.51\pm0.05$) for the low $T_x/E(z)$ clusters and ($\alpha=0.98\pm0.28$, $\beta=0.40\pm0.04$) for the clusters with no $T_x$ information.  This consistency is compatible with a GDR, and GDR evolution that do not depend on cluster mass over the range of mass and redshift probed by our sample.

\begin{figure*}
\begin{flushleft}
\begin{minipage}[b]{0.48\linewidth}
    \includegraphics[width=3.4in]{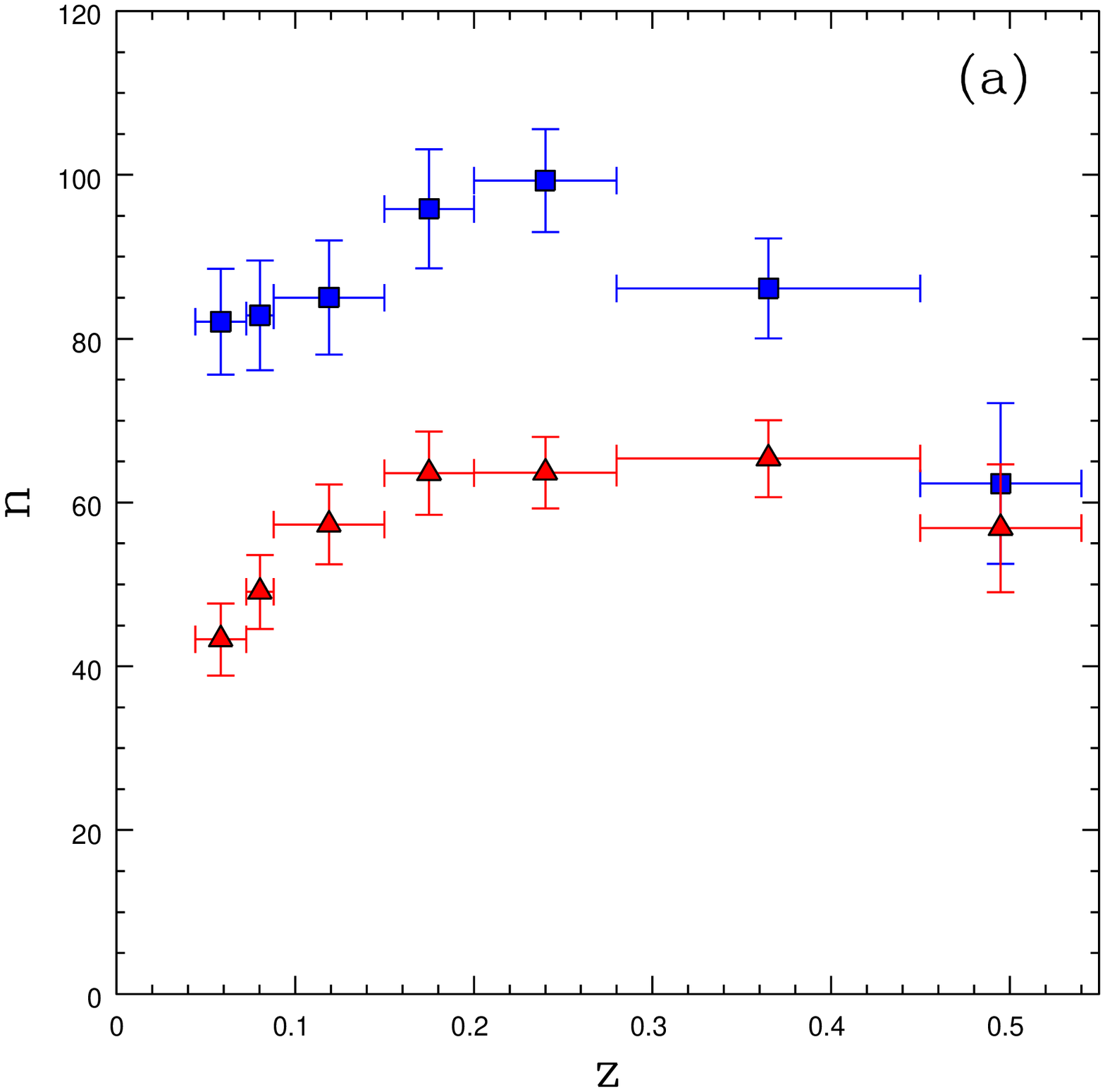}
\end{minipage}
\hspace{0.5cm}
\begin{minipage}[b]{0.48\linewidth}
    \includegraphics[width=3.4in]{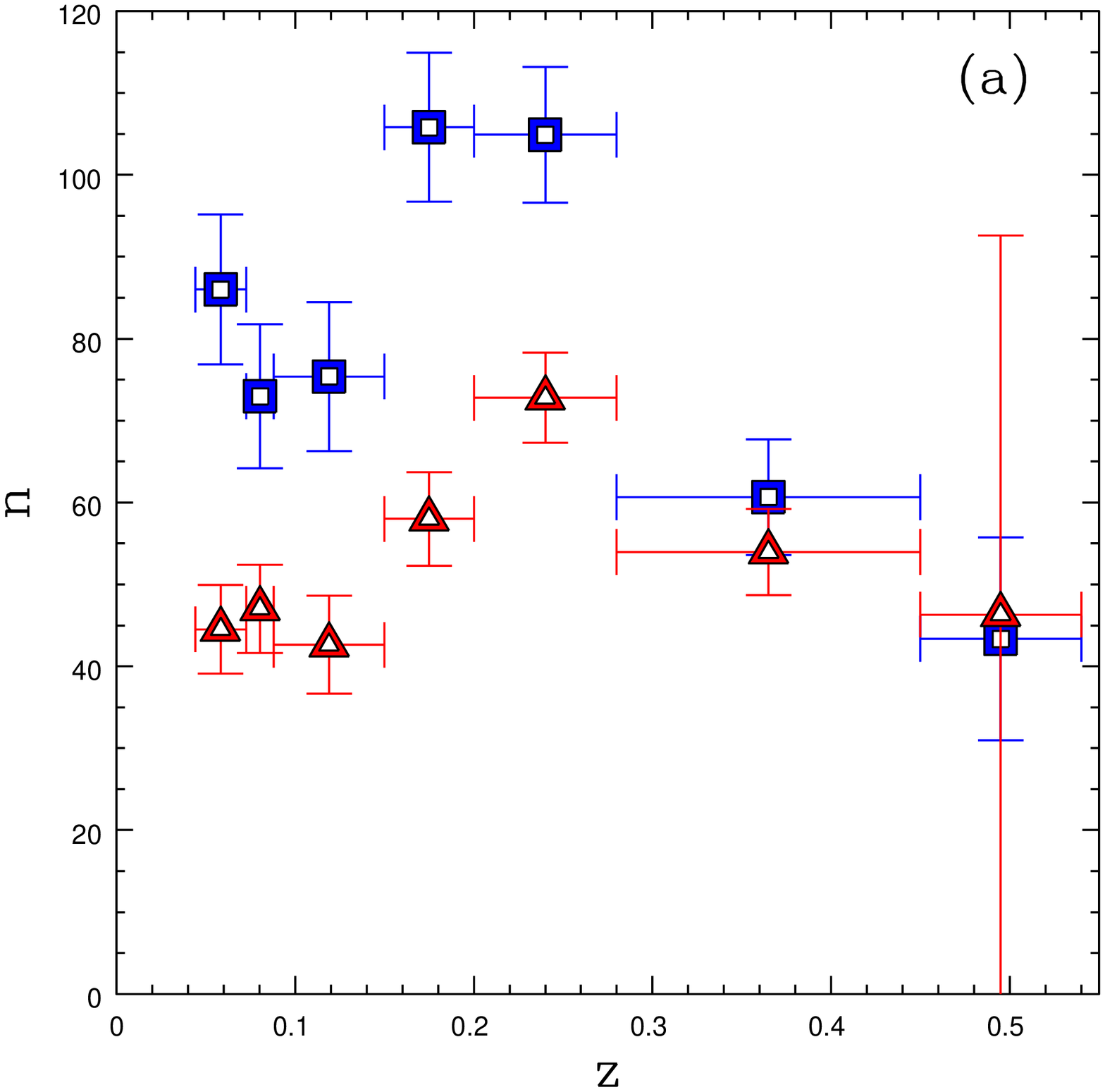}
\end{minipage}
\caption{The evolution of the number of dwarfs (blue squares) and giants (red triangles) after correcting to a common $T_x/E(z) = 6$ keV.  Vertical error bars show the uncertainty on the error-weighted bin centroid and horizontal error bars indicate the width of the bins.  The data in the left panel \textbf{(a)} shows the number of galaxies in a projected radius $r < 750$ kpc (default region) while right panel \textbf{(b)} shows the number of galaxies in a region $750$ kpc $< r < 1500$ kpc.}
 \label{fig:ngd}
\end{flushleft}
\end{figure*}

A subset of 27 clusters in our sample are part of a more detailed campaign to measure the masses using a combined X-ray + weak lensing method.  We compare the GDR directly to the weak-lensing masses (taken from Hoekstra et al. in prep.) but find no evidence for a correlation.  This is consistent with the result obtained using $T_x/E(z)$ as a mass proxy.  From the X-ray portion of the analysis we obtain central entropy values ($S_0$) for the subset (taken from Mahdavi et al. in prep.).  The value of $S_0$ is often used to discriminate between cool-core and non-cool-core clusters, so it is interesting to examine if/how the GDR varies with $S_0$.  We find no correlation between these two parameters, however, indicating that the GDR is similar between cool-core and non-cool-core clusters.  Lastly, we consider the GDR as a function of the projected distance between the BCG and the peak of the cluster X-ray emission, an indicator of the dynamical state of the cluster.  Using values from Bildfell et al. (2008) we find no evidence for a correlation between GDR and BCG to X-ray peak distance.  These results are all consistent with, and expected for a GDR vs $z$ relation that shows no intrinsic scatter.

\section{Driver of Evolution} \label{driver}

The physical mechanism(s) responsible for the evolution in the GDR for $0.05<z<0.55$ is not evident.  The two main processes affecting red sequence galaxies often discussed in the literature are dry mergers, which reduces the number of galaxies on the red sequence and late-time quenching of star formation, which increases the number of galaxies on the red sequence.  Both of these mechanisms are taking place in and around clusters but the extent to which they operate and how their behavior may change with redshift remains elusive.  Bell et al. (2004) and Faber et al. (2007) argue that both quenching and dry merging are required to explain observations of the red sequence buildup since $z\sim1$.  Observational evidence for dry mergers of early type galaxies at $0 < z < 0.8$ is presented in van Dokkum (2005) and Tran et al. (2005).  On the other hand, work by Cimatti et al. (2006) and Scarlata et al. (2007) argue that the buildup of the red sequence since $z\sim0.7$ and $z\sim1$, respectively, can be explained by the late-time quenching of star formation alone.  With the exception of Tran et al. (2005) however, these studies are based on samples containing galaxies in the field and not exclusively on cluster galaxies where the high-density local environment may be important.  To investigate this further, we examine the evolution of the dwarf and giant populations separately.

We showed in \S \ref{clusterprops} that the GDR is not strongly dependent on cluster mass (or $T_x$/E(z)).  When considered independently however, $n_g$ and $n_d$ are both expected to scale with cluster mass; similar to the known scaling between richness and $T_x$ (Yee \& Ellingson 2003).  Such a scaling is implied by the existence of a mass-richness relation.  This effect must be accounted for before we examine the evolution of the dwarf and giant populations separately.  We show in Figure \ref{fig:ngd_tx} the dependence of galaxy number $n_g$ and $n_d$ on $T_x/E(z)$.  The data are fit using a linear relation between number density and $T_x/E(z)$.  To minimize the effect of the correlation between $T_x/E(z)$ and $z$ imposed by our sample selection, we only fit clusters in the range $0.1 < z < 0.3$.  We find a best fit slope $\alpha_g=8.4$ keV$^{-1}$ for the giants and $\alpha_d=12.1$ keV$^{-1}$ for the dwarfs.  Based on this temperature dependence we correct all $n_g$ and $n_d$ values to a common $T_x/E(z) = 6$ keV.
  
We plot the corrected values as a function of redshift in Figure \ref{fig:ngd}a.  The data are combined using a constant number of clusters in each bin (horizontal error bars indicate bin limits).  Both the number of dwarfs and the number of giants increase from $z\sim0.55$ up to $z\sim0.3$, with the dwarfs showing the more dramatic evolution.  From $z\sim0.3$ to $z\sim 0.2$ the numbers of dwarfs and giants remain relatively constant.  Below $z\sim0.2$ the number of dwarfs remains constant but the number of giants begins to decrease significantly.  The results suggest that the GDR evolution at $z < 0.2$ is dominated by mergers between giant galaxies.  When considered with GDR results at higher redshift, the data support a transition in the dominant mechanism governing GDR evolution, which is often attributed to quenching in dwarfs.  This interpretation could also explain the findings of Lu et al. (2009), who measure a rapid DGR evolution at $0 < z < 0.2$ and then a flattening of the relation out to $z\sim0.4$.  Though not discussed in their work, Figure 19 of Lu et al. (2009) clearly shows a reduction in the number of giants from $z\sim0.2$ to $z\sim 0.1$.

Given the evidence supporting merger-driven evolution at low redshift, it is natural to ask whether this result is compatible with theoretical predictions.  The orbit of a galaxy about the cluster centre will decay with time due to dynamical friction.  The relevant timescale for the orbital decay of an individual galaxy can be expressed as:
 
 \begin{eqnarray}
  t_{dyn} & \approx & 6 \times 10^9 \textrm{yr} \left(\frac{\sigma_r}{1000 \textrm{km/s}}\right) \left(\frac{r_c}{0.25 \textrm{Mpc}}\right)^2 \nonumber \\
  & & \times \left(\frac{m}{m^*}\right)^{-1} \left[ \frac{(M/L_V)_{gal}}{10M_\odot/L_\odot}  \right]^{-1}
 \label{eqn:tdyn}
 \end{eqnarray}

\noindent with line-of-sight velocity dispersion $\sigma_r$, cluster core radius $r_c$, galaxy mass $m$ and the galaxy's V-band mass-to-light ratio $(M/L_V)$.  For a giant galaxy with $m=m^*$ and a $M/L_V=10 M_\odot/L_\odot$ orbiting near the centre of a cluster where $\sigma_r\sim600$ km/s and $r_c=0.25$ Mpc the expected timescale is $t_{dyn}\approx 3.6$ Gyr.  Over the redshift range of our data ($0.05 < z < 0.55$) the total change in cosmic time is $\Delta t_{age}= 4.7$ Gyr, which we use as an estimate of the elapsed interaction time.  This is considered to be a conservative estimate because these clusters are likely assembled at much earlier epochs.  Since $t_{dyn} < \Delta t_{age}$ we should indeed expect to see the signature of the most massive galaxies merging together on these timescales.  This prediction is in good qualitative agreement with the decrease in the number of giants galaxies at low redshift in Figure \ref{fig:ngd}a.  Furthermore, quenching is expected to become less important at late times due to the declining specific star formation rate (at fixed galaxy mass) in newly accreted field galaxies (eg. Juneau et al. 2005).
 
If the GDR evolution seen at low redshift is driven by orbital decay and mergers in the giant population then we might expect to see less of this effect taking place in the cluster outskirts, where the density of the ambient medium is lower.  To test this we plot in Figure \ref{fig:ngd}b the number of giants and dwarfs (adjusted to $T_x/E(z) = 6$ keV) in an annulus with the range of projected radii 750 kpc $< r <$ 1500 kpc.  The number of giants rises from $z\sim0.5$ and peaks near $z\sim0.25$ declining and flattening below $z<0.15$.  The number of dwarfs generally increases from $z\sim0.5$ to $z\sim0.05$ but the distribution is somewhat noisier than similar data at smaller projected cluster-centric distance (Figure \ref{fig:ngd}a).  The triggering of star formation as galaxies fall into the cluster, pre-processing of galaxies in groups, and the increased contamination by interloper galaxies may all affect the number counts in the cluster outskirts and complicate their interpretation considerably.

It is difficult to explain all of the features of Figure \ref{fig:ngd} in the context of mergers and quenching alone.  The rate of newly infalling cluster galaxies, as it depends on redshift and galaxy mass, is also likely to be a key ingredient.  The merger tree models of de Lucia et al. (2011) indicate that up to $\sim50$\% of the massive galaxies ($\log M_*/M_\odot > 10$) that fall into clusters do so after $z=1$.  Taking this into consideration, it may be more appropriate to describe the GDR evolution at low redshift as competition between the galaxy infall rate and the effects of dynamical friction, with quenching as a secondary effect.

\begin{figure}
    \centering
    \includegraphics[width=3.2in]{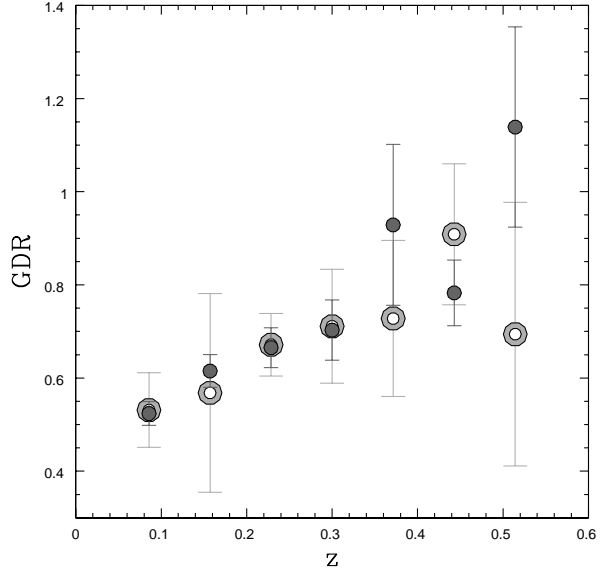}
    \caption{The evolution of the GDR in two distinct regions of projected-radius.  Filled symbols represent the GDR inside a projected radius limit $r < 750$ kpc (default region), while open symbols represent the GDR in the range $ 750$ kpc $< r < 1500$ kpc.}
    \label{fig:gdr_an}
 \end{figure}

For a final check we compare in Figure \ref{fig:gdr_an} the GDR evolution in the cluster outskirts (750 kpc $< r <$ 1500 kpc) with that in the cluster centre ($r < 750$ kpc).  We note however, that the data are slightly incomplete at the lowest redshifts where the field of view of MegaCam extends only to 1490 kpc from the BCG (see \S \ref{data}).  The results indicate that the GDR evolution, as measured in the cluster interior and outskirts, are in good agreement with each other.  This suggests that the physical mechanism(s) responsible for driving the GDR evolution operate both at large and small cluster-centric distance.  This may simply reflect the fact that clusters are assembled from groups and that infalling galaxies are likely to have already undergone some processing in high-density environments.

\subsection{Integrated Stellar Masses}

To test the interpretation of a merger-driven GDR evolution at $z < 0.2$, we measure the integrated stellar mass of giants ($M_*^g$) and dwarfs ($M_*^d$), separately, as a function of $z$.  We rearrange equation \ref{eqn:obscuts} to obtain $(K+E)$-corrected $M_{r'}$ values for individual galaxies which are then converted to $r'$-band luminosities $L_{r}$ assuming an $r'$-band absolute magnitude for the sun $M_{r,\odot}=4.65$ \footnote{http://mips.as.arizona.edu/~cnaw/sun.html}.  The luminosities of interloper galaxies are calculated in a similar way using the CFHTLS Deep background populations described in \S \ref{bgcolcor}.  The integrated dwarf or giant luminosity is obtained for each cluster by summing $L_r$ values for a given population and then subtracting the appropriate background.  We also apply the colour error correction described in \S \ref{bgcolcor}.  To convert the luminosity to a stellar mass we assume a universal stellar mass-to-light ratio in the $r'$-band of $M/L_r = 1.48$, which is obtained from the scaling relation of Zibetti et al. (2009) for elliptical galaxies with $(g-r)=0.61$.  We note however, that the exact normalization is unimportant for our purpose, as we are primarily interested in the relative change in stellar mass.

\begin{figure}
    \centering
    \includegraphics[width=3.2in]{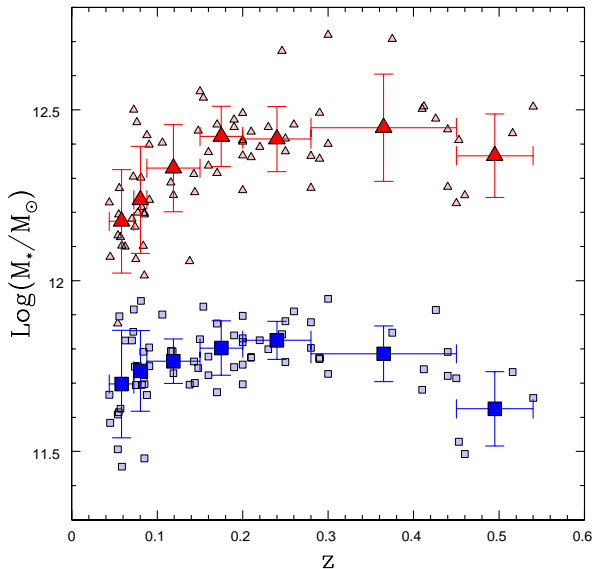}
    \caption{Integrated masses of galaxies associated with the dwarf population (blue squares) and giant population (red triangles) as a function of redshift.  The data have been scaled to a common cluster Xray temperature $T_x = 6/E(z)$ keV.  Individual clusters are shown as smaller, lighter symbols, while binned data is shown as larger, darker symbols.  Error bars on binned data indicate the bin width (horizontal) and the standard deviation within the bin (vertical).  The mass in dwarf galaxies is roughly constant, while the mass in giants is decreasing below $z < 0.2$.}
    \label{fig:mgd}
 \end{figure}

To make a fairer comparison of the masses, we fit the $\log{M_*}$ versus $T_x/E(z)$ relation (for $0.1 < z < 0.3$) and correct to a common cluster temperature of $T_x/E(z)=6$ keV.  This procedure is analogous to that in \S \ref{driver} regarding the number densities.  We show in Figure \ref{fig:mgd} the $T_x/E(z)$-corrected values of $M_*^g$ and $M_*^d$ as a function of redshift.  The data reveal that the mean stellar mass in dwarf galaxies ($\langle M_*^d \rangle$) increases from $z\sim0.55$ to $z\sim0.25$ and remains roughly constant for $z<0.25$.  In contrast, the mean stellar mass in giants ($\langle M_*^g \rangle$) is roughly constant from $z\sim0.55$ to $z\sim0.2$ and then declines at $z<0.2$.

In \S \ref{driver} we propose that the decrease in the number of giants at $z < 0.2$ can be explained by a significant number of giant-giant mergers (merger scenario).  If this is the case then one might expect $\langle M_*^g \rangle$ to remain constant over this redshift range.  However, there are several reasons why mergers may act to deplete the integrated stellar mass.  Tidal features created in mergers are believed to contribute to the production of intra-cluster stars (ICS), which are not gravitationally bound to any particular galaxy and make up $\sim10\% - 30\%$ of the stellar mass in clusters (Gal-Yam et al. 2003; Gonzalez et al. 2005; Zibetti et al. 2005; Krick \& Berstein 2007, Sand et al. 2011).  Furthermore, the existence of a tight relationship between effective radius and mean effective surface brightness (Kormendy 1977) implies that as elliptical galaxies grow, they become increasingly diffuse.  Consequently, the fraction of a galaxy's stellar mass that is below our detection threshold may be a function of the number of mergers that galaxy has undergone.

We see from Figure \ref{fig:ngd} that there is a 30$\pm$13\% reduction in the number density of giants from $z\sim0.2$ to $z\sim0.05$.  The corresponding reduction in $\langle M_*^g \rangle$ over this period is 38$\pm$14\%.  Assuming the merger scenario is correct, a sizeable fraction of the stellar mass involved in a giant-giant mergers, up to 100\% in some cases, must be shifted to the ICS or drop below our detection threshold.  As a caveat to these statistics, Gonzalez et al. (2007) finds that the ICS fraction scales inversely with the velocity dispersion of the cluster.  If ICS production is much more efficient in the lowest mass clusters then this effect may be exacerbating the observed drop in $\langle M_*^g \rangle$ at $z < 0.2$, where our sample is dominated by clusters with low $T_x$.  To test this we repeat our measurement using only those clusters with $T_x/E(z) > 6$ keV.  With this selection, the drop in $\langle M_*^g \rangle$ at $z<0.2$ is less pronounced and $\langle M_*^g \rangle$ is consistent with a constant value over all $z$.  These results are suggestive but their interpretation is limited by the large standard deviation within each bin.

Because of the above mentioned considerations we are unable to confirm/deny the merger scenario with the integrated stellar masses alone.  To make progress requires measurement of the ICS fraction for all of the clusters in our sample and better constraints on the redshift-dependance of the ICS fraction, both of which are outside the scope of this paper.

\section{Conclusions} \label{conclusions}

We analyzed a large sample of 97 galaxy clusters observed in $g'$ and $r'$ with CFHT-MegaCam.  After carefully accounting for several observational biases and statistically removing interloper galaxies using $g'$ and $r'$ galaxy catalogs generated from the CFHTLS Deep fields as a control, we measured the GDR in galaxy clusters over the range $0.05 < z < 0.55$.  We find several interesting results, the most important of which we summarize below.

We detect evolution in the GDR over the redshift range $0.05 < z < 0.55$, which can be parametrized by a simple linear relation with $z$ ($GDR=\alpha z + \beta$).  The best fit parameters are $\alpha = 0.88\pm0.15$ and $\beta=0.44\pm0.03$ with an estimate for the intrinsic scatter in the relation that is consistent with zero.  Neglecting the uncertainty introduced by the presence of large-scale structure leads to a perceived intrinsic scatter of $\sigma_{int}=0.088\pm0.017$, which is consistent with previous estimates (Andreon et al. 2008).

Varying the method of selecting dwarf and giant galaxies by altering the BCG-centric distance threshold, colour selection criteria or $K$-correction prescription does not change the GDR significantly.  So long as one makes reasonable choices about the selection criteria and $K$-correction, then varying these parameters cannot hide the signature of an evolving GDR.

The GDR values found here agree well with the those from the literature that are derived directly, solely from measurements of the number of luminous and faint galaxies.  However, our GDR results are not in agreement with those that are inferred from parametric fits to the luminosity function (eg. Andreon 2008, Crawford et al. 2009).

The GDR does not correlate with cluster mass estimates via X-ray temperature or weak-lensing.  Similarly, we find no evidence for a correlation between GDR and the central entropy of the ICM or the dynamical state of the cluster (BCG to X-ray peak offset).  This is further validated by the good agreement between our GDR values and those found in studies based on lower mass clusters (eg. Gilbank et al. 2008, Capozzi et al. 2010).  These result agree with our finding that the intrinsic scatter in GDR vs $z$ is consistent with zero.  When considered individually however, both the number of dwarfs and the number of giants are strongly correlated with $T_x/E(z)$, which is expected from the known correlation between richness and $T_x$ (Yee \& Ellingson 2003).

Separate inspection of the evolving number of dwarfs and giants, after correcting to a common $T_x/E(z)$, suggests a change in the primary physical mechanism responsible for the evolution in the GDR over the redshift range $0.05 < z < 0.55$.  From $z\sim0.55$ to $z\sim0.25$ we observe an overall increase in the number of both dwarfs and giants, with a more rapid increase among the dwarfs.  Below $z\sim0.2$ however, the number of dwarfs is roughly constant, while the number of giants decreases significantly.  To explain the transition at $z\sim0.2$, we argue for a significant number of mergers within the giant population, caused by orbital decay due to dynamical friction.  This is supported by calculations of the expected orbital decay time for giant galaxies, which yield timescales smaller than the span in look-back time of our sample.  In this scenario, the GDR is governed by the competition between mergers, which become dominant at $z<0.2$, versus the processes that act to increase the relative number of dwarf galaxies on the red sequence, such as the late-time quenching of star formation and the redshift dependence of the accreted-galaxy mass function.  We consider the latter to be particularly important given the results of de Lucia et al. (2011) indicating that up to 50\% of $\log M_*/M_\odot > 10$ galaxies fall into clusters at $z<1.0$.  A GDR evolution that is driven by a reduction in the number of giants at $z< 0.2$ could also explain why Crawford et al. (2009) find no evolution in the faint-end slope of the luminosity function at low redshift.

To test the merger-driven GDR scenario, we measure the integrated stellar mass in giants and dwarfs as a function of redshift.  We find that $\langle M_*^d \rangle$ increases from $z\sim0.55$ to $z\sim0.25$ and remains roughly constant for $z<0.25$, while $\langle M_*^g \rangle$ is roughly constant from $z\sim0.55$ to $z\sim0.2$ and then declines at $z<0.2$.  Because of the large scatter present in the values of $M_*^g$ at fixed $z$ and the uncertainty regarding the production of ICS, it is not possible to confirm nor deny the merger scenario with this measurement alone.  To make progress requires measurement of the ICS fractions of clusters in our sample and better constraints on the evolution of the ICS.

The results presented in this paper clearly indicate an evolving GDR in our sample of 97 galaxy clusters at intermediate redshift.  We are cautious, however, not to over-interpret the data in light of the fact that the sample is incomplete.  Correlation between the GDR and global cluster properties, along with a covariance between said properties and $z$ in our sample, could be misinterpreted as an enhancement of the GDR evolution.  The data presented in \S \ref{clusterprops}, however, along with the good agreement with the literature values over a range of sample selection properties, and an intrinsic scatter that is consistent with zero, all indicate that this effect is negligible.  A future study of the GDR using a sample that represents an evolutionary sequence of clusters (c.f. Hart et al. 2011), along with spectroscopic redshifts for identifying cluster members would be useful to confirm/deny these findings and allow for a clearer interpretation that is unaffected by potential selection biases.  Numerical models of cluster formation that focus on the competition between mergers of existing cluster members versus the recent infall of new cluster galaxies, as it depends on galaxy mass and redshift, would be valuable for assessing the proposed mechanisms for driving GDR evolution.

\section{Acknowledgments}

CB would like to acknowledge some very helpful discussions with Gabriella de Lucia, Michael Balogh, Simone Weinmann, Ting Lu, Stephen Gwyn and Diego Capozzi.  CB and HH acknowledge support from Marie Curie International Reintegration Grant (number 230924).  HH also acknowledges support from a Vidi grant from the Netherlands Organization for Scientific Research (NWO, grant number 639.042.814).  AB and CB would like to acknowledge support from NSERC through the Discovery Grant program. They would also like to thank John Criswick for support through his generous gifts to the AB and to the Department of Physics and Astronomy, University of Victoria.  This research has made use of the X-Rays Clusters Database (BAX) which is operated by the Laboratoire d'Astrophysique de Tarbes-Toulouse (LATT), under contract with the Centre National d'Etudes Spatiales (CNES).  Finally, CB, HH and AB thank the Lorentz centre in Leiden for hospitality and support  over the course of the Spring 2010 and Summer 2011 Workshops during which time a significant fraction of the research reported here was carried out.

\section{References}
Andreon S., 2008, MNRAS, 286, 1045\\
Barkhouse W., Yee H., Lopez-Cruz O., 2007, ApJ, 671, 1471\\
Beers T., Flynn K., Gebhardt K., 1990, AJ, 100, 32\\
Bell E. et al., 2004, ApJ, 608, 752\\
Bertin E, Arnouts S., 1996,  A\&AS, 317, 393\\
Bertin E., Mellier Y., Radovich M., Missionnier G., Didelon P., Morin B., 2002, ASPC, 281, 228\\
Bildfell C., Hoekstra H., Babul A., Mahdavi A., 2008, MNRAS, 389, 1637\\
Bundy K., Ellis R., Coselice C., 2005, ApJ, 625, 621\\
Butcher H., Oemler A., 1984, ApJ, 285, 426\\
Bower R., Lucey J., Ellis R., 1992, MNRAS, 254, 601\\
Bower R., Benson A., Malbon R., Helly J., Frenk C., Baugh C., Cole S., Lacey C., 2006, MNRAS, 370, 645\\
Capozzi D., Collins C., Stott J., 2010, MNRAS, 403, 1274\\
Cardelli J., Clayton G., Mathis J., 1989, ApJ, 345, 245\\
Cimatti A., Daddi E., Renzini A., 2006, A\&A, 453L, 29\\
Crawford S., Bershady M., Hoessel J., 2009, ApJ, 690, 1158\\
De Lucia G. et al., 2007, MNRAS, 374, 809\\
De Lucia G., Blaizot J., 2007, MNRAS, 375, 2\\
De Lucia G., Weinmann S., Poggianti B., Aragon-Salamanca A., Zaritsky D., 2011, arXiv:1111.6590v1\\
Evrard A., Metzler C., Navarro J., 1996, ApJ, 469, 494\\
Faber S. et al., 2007, ApJ, 665, 265\\
Fukugita M., Ichikawa T., Gunn J., Doi M., Shimasaku K., Schneider D., 1996, AJ, 111, 1748\\
Gal-Yam A., Maoz D., Guhathakurta P., Filippenko A., 2003, AJ, 125, 1087\\
Gilbank D., Yee H., Ellingson E., Gladders M., Loh Y., Barrientos L., Barkhouse W., 2008, ApJ, 673, 742\\
Gilbank D., Balogh M., 2008, MNRAS, 385L,116\\
Gonzalez A., Zabludoff A., Zaritsky D., 2005, ApJ, 618, 195\\
Gwyn S., 2008, PASP, 120, 212\\
Hansen S., Sheldon E., Wechsler R., Koester B., 2009, ApJ, 699, 1333\\
Hart Q., Stocke J., Evrard A., Ellingson E., Barkhouse W., 2011, ApJ, 740, 59\\
Hoekstra H., 2007, MNRAS, 379, 317\\
Horner D., 2001, PhDT, 88\\
Juneau S. et al., 2005, ApJ, 619L, 135\\
Kodama T., Arimoto N., 1997, A\&A, 320, 41\\
Kormendy J., 1977, ApJ, 218, 333\\
Krick J., Bernstein R., 2007, AJ, 134, 466\\
Lu T., Gilbank D., Balogh M., Bognat A., 2009, MNRAS. 399, 1858\\
Markwardt C., 2009, ASPC, 411, 251\\
Mei S. et al., 2006, ApJ, 644, 759\\
Pimbblet K., Smail I., Kodama T., Couch W., Edge A., Zabludoff A., O'Hely E., 2002, MNRAS, 331, 333\\
Rozo E. et al., 2009, ApJ, 699, 768\\
Scarlata C. et al., 2007, ApJS, 172, 494\\
Sand D. et al., 2011, 729, 142\\
Sand D. et al., 2012, ApJ, 746, 163\\
Sarazin C., 1998, X-ray Emission from Clusters of Galaxies, Cambridge University Press\\
Schlegel D., Finkbeiner D., Davis M., 1998, ApJ, 500, 525\\
Stott J., Smail I., Edge A., Ebeling H., Smith G., Kneib J., Pimbblet K., 2007, ApJ, 661, 95\\
Tran K., van Dokkum P., Franx M., Illingworth G., Kelson D., Schreiber N., 2005, ApJ, 627\\
van Dokkum P., 2005, AJ, 130, 2647\\
Visvanathan N., 1978, A\&A, 67, L17\\
Williams M., Bureau M., Cappellari M., 2010, MNRAS, 409, 1330\\
Yee H., Ellingson E., 2003, ApJ, 585, 215\\
Zibetti S., White S., Schneider D., Brinkmann J., 2005, MNRAS, 358, 949\\
Zibetti S., Charlot S., Rix H., MNRAS, 400, 1181\\

\begin{appendix}

\section{Red Sequence Fitting} \label{redseqappendix}

As mentioned in \S \ref{redseq} we use the biweight estimator of Beers et al. (1990) to fit the colour-magnitude relation (CMR) for the galaxy distribution inside a projected radius of 500 kpc from the BCG.  Lacking spectroscopic redshifts for the cluster galaxies, we begin the fitting procedure by estimating a region in the CMD that is likely to be minimally contaminated by interloper galaxies.  To the points lying inside this region we then apply a red sequence fitting procedure similar to that used by Pimbblet et al. (2002).  We first determine the probability that a galaxy belongs to the field population using equation \ref{eqn:pfield}:

\begin{equation}
P_{f}(m_{r'},g'-r')=\frac{\Sigma_{f}(m_{r'}, g'-r')}{\Sigma_{f}(m_{r'}, g'-r')+\Sigma_{c}(m_{r'}, g'-r')}
\label{eqn:pfield}
\end{equation}

\noindent where $\Sigma_{f}$ is the surface density of field galaxies, $\Sigma_{c}$ is the surface density of cluster galaxies and both are functions of $m_{r'}$ and $(g'-r')$.  The surface densities are estimated by generating a 2-d histogram of galaxies in colour-magnitude space using a bin height and width of $\Delta (g'-r')=0.078$ and $\Delta m_{r'}=0.56$.  The denominator in equation \ref{eqn:pfield} is calculated using galaxies within a projected radius of 500 kpc from the BCG where there is a mix of the cluster and field populations, while the numerator is calculated using galaxies that are outside a projected radius of 1.5 Mpc where the sample should be more dominated by the field population.  We chose the 1.5 Mpc limit in order to be as far away from the BCG as possible while maintaining a large enough field-sampling region at the low redshift end of the survey $z \sim 0.05$.  At this redshift roughly 45\% of the image is available for estimating the field population.  With each galaxy assigned a probability of being a field member ($P_{field}$) and a probability of being a cluster member ($1-P_{field}$) we then fit the red sequence to the points inside the selection box.  We randomly reject points inside the initial selection box based on their probability of being members of the field and then perform an unweighted, linear least-squares fit on the remaining data.  This procedure is repeated 100 times.  We take the median values of these 100 solutions as the best fitting slope and intercept parameters with their errors determined by the standard deviation.

\begin{figure*}
    \centering
    \includegraphics[width=7.5in]{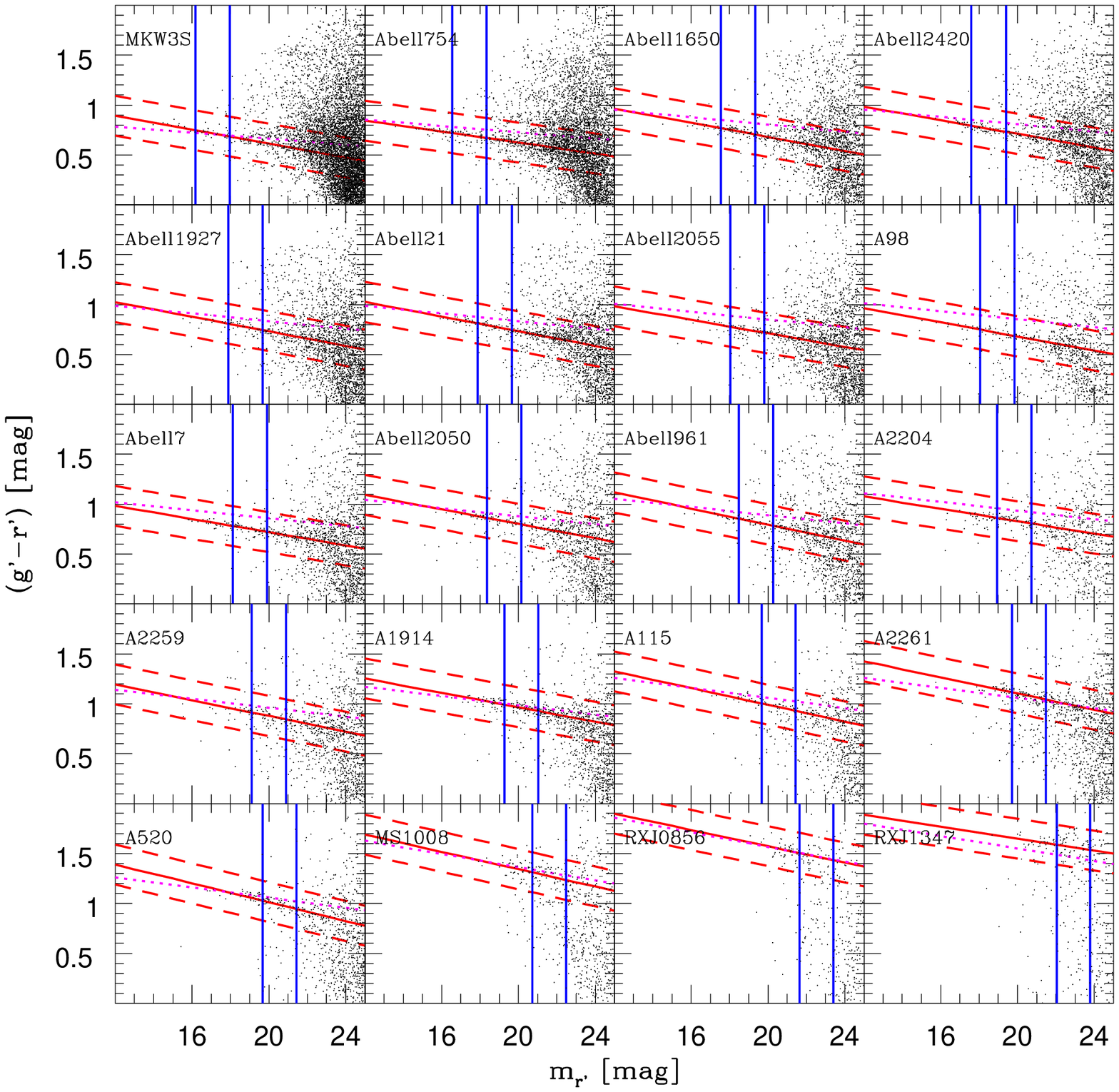}
    \caption{Illustration of the red sequence fitting procedure for 20 clusters selected randomly from our sample.  They are arranged in increasing redshift from the top left.  The solid red line shows the CMR according to a linear fit to the data while the dotted magenta line shows the CMR defined by a stellar population that was created in a burst at $z=3$ and aged passively (see text).  The dashed red lines show the colour offset $\Delta (g'-r')=\pm 0.2$ used for selecting red sequence members.  The vertical blue lines show the limits to classify dwarfs and everything brighter than this is classified as a giant.}
    \label{fig:redseq_b}
\end{figure*}

The results of the red sequence fitting for a subset of 20 randomly selected clusters are plotted in figure \ref{fig:redseq_b}.  In each panel we show, for a given cluster, the best fit red sequence, a model red sequence for comparison and the dwarf/giant magnitude limits at the corresponding redshift.  The model red sequence is constructed assuming that the correlation between galaxy colour and magnitude is driven solely by an increasing metallicity with stellar mass.  We use the Charlot \& Bruzual (2007) synthetic stellar population code  (hereafter CB07) with a Salpeter initial mass function and a single burst of star formation at $z=3$ to compute synthetic galaxy colours over a range of metallicities.  Each model CMR also contains a small contribution ($3\%$ by mass) of ultra metal-poor stars, which Maraston et al. (2009) find is necessary to reproduce the $(g'-r')$ colour of early type galaxies at similar redshifts.  Without this component the models are up to 0.1 magnitudes too blue.  The model is then calibrated to reproduce the magnitude vs metallicity relation Kodama \& Arimoto (1997).  It can be seen from Figure \ref{fig:redseq_b} that the model CMR is in qualitative agreement with the observations, including their evolution.  The model CMR slope however is typically shallower than the best fit CMR and this difference in slope can vary appreciably from cluster to cluster.  This is likely a consequence of the over-simplicity of calibration to a fixed magnitude metallicity relation and the use of a single stellar population.  We take the model as illustrative only and for the purposes of classifying giants and dwarfs we use the CMR as defined by the fit.

\end{appendix}

\end{document}